\documentclass[%
 reprint,
 amsmath,amssymb,
 onecolumn,
prb,
]{revtex4-2}

\usepackage{graphicx}
\usepackage{dcolumn}
\usepackage{bm}

\usepackage{caption}
\usepackage{subcaption}
\usepackage{float}

\begin{document}

\preprint{}

\title{Low-energy scattering parameters:  A theoretical derivation of the 
effective range and scattering length for arbitrary angular momentum}

\author{Jordi Pera}
\author{Jordi Boronat}%
\affiliation{%
Departament de Física, Campus Nord B4-B5, Universitat Politècnica de Catalunya, 
E-08034 Barcelona, Spain
}%


\begin{abstract}
The most important parameters in the study of low-energy scattering are the 
$s$-wave and $p$-wave scattering lengths, and the $s$-wave 
effective range. We solve the scattering problem and find 
two useful formulas for the scattering length and the effective range 
 for any angular momentum, as long as the Wigner threshold law holds. 
Using that formalism, we obtain a set of useful formulas for the angular-momentum scattering parameters of four different model potentials: hard-sphere, soft-sphere, spherical well, and  well-barrier potentials.
The behavior of the scattering parameters close to Feshbach 
resonances is also analyzed. Our derivations can be useful as hands-on 
activities for learning scattering theory.
\end{abstract}

\maketitle

\section{Introduction}

The natural way to learn about the forces acting between particles is 
to observe their mutual interaction. Most of what we know 
about the micro-world has been established by means of collision processes. 
In physics, a well-studied collision process is 
the scattering of an incident particle from a stationary target. A free particle (or rather 
a beam of such particles) with known characteristics collides with a target 
particle, interacts with it, and scatters into a modified free state. 
One then measures the energy, angular distribution, and other characteristics of the scattered beam and infers from them the nature and 
strength of the forces which, during the collision, acted between the projectile 
and the target \cite{roman}.
In the theoretical analysis of an scattering process, the asymptotic free states of the scattered particles are the most relevant. Therefore, it is not necessary to give an interpretation, or even to have a detailed knowledge, of the state vector of the entire system when the particles are close and interact strongly. This remark is important in connection with quantum field theory, where the interpretation of the states of strongly interacting fields is extremely difficult, if possible at all. However, it is relatively easy to discuss the asymptotic free states \cite{roman}.

Low-energy scattering parameters are fundamental in the theoretical description of interacting many-body systems. By low-energy we mean that the energy of the particles is much smaller than the typical excitation energy of the target.
The equations of state of dilute Bose and Fermi gases depend on these 
parameters, which account for the full inter-atomic potential if the density is 
low enough \cite{scaguardiola,Bishop}, that is, the range of the interaction is much smaller than the interparticle distance $\sim\rho^{-1/3}$. In particular, the so called universal 
regime is the one in which the interaction is fully described by a single 
parameter, the $s$-wave scattering length. In the universal regime, any 
potential with the same scattering length gives the same energy, independently 
of its particular shape. However, when the density of the system is increased, effects of the shape of the potential arise and we need to consider more scattering parameters, such as the $s$-wave effective range. In some recent studies, these effects have been reported \cite{viktor,raul}. In this work, we find both scattering lengths and effective ranges.
This leads to a kind of 
inverse problem; that is, we know a few scattering parameters of the 
physical system and the goal is to find a model potential with the same 
scattering parameters. Then, we can use this model potential 
to study the many-body properties of the system.
Obviously, the solution is not unique 
and model potentials as simple 
as possible are sought for this purpose.

This paper presents a well-detailed method that leads to integral expressions that allow the calculation of the scattering parameters for any angular momentum of the system, as long as the Wigner threshold law holds, that is, potentials decaying at large r faster than $1/r^n$, with $n>2l+3$ for each partial wave. This procedure is not only useful for researchers who want to know the value of the scattering parameters, but also for students learning scattering theory. The scattering problem has been widely studied and there are many procedures leading to similar expressions. H. Bethe found the expressions for $l=0$ scattering \cite{bethe}, and the generalization for any angular momentum $l$ was done by L. B. Madsen in 2002 \cite{lars}.

We show that the low-energy scattering parameters can be accurately calculated 
by an integral equation method, allowing their determination for any angular 
momentum. Using this result, we discuss some properties of the scattering 
parameters around a Feshbach resonance~\cite{chin}. The Feshbach resonance 
occurs when the energy of a scattering state is very close to the energy of a 
bound state and the scattering length diverges. When the divergence is crossed, 
a bound state appears. We apply the formulas for getting analytical results of the angular momentum scattering lengths and effective ranges for the following potentials: hard-sphere, soft-sphere, spherical well, and well-barrier potential.

The rest of the paper is organized as follows. Sec. II introduces basic 
relations for low-energy scattering. Sec. III contains the derivation of the 
formalism for the calculation of the scattering length and effective range for 
any angular momentum $l$. In Sec. IV, the behavior of the parameters and the 
Feshbach resonance are addressed. Sec. V is organized around the four considered 
potentials: Subsec. A corresponds to the hard-sphere potential, Subsec. B to the 
soft-sphere potential, Subsec. C to the spherical well potential and Subsec. D 
to the well-barrier potential. Finally, the main conclusions are discussed in 
Sec. VI.

\section{Low-energy scattering}
\label{secII}
The study of low-energy scattering events requires the characterization of scattering parameters such as the scattering lengths and ranges. One convenient approach to find these parameters is to study the phase shift due to the effect of a 
potential. We need to have in mind that the phase shift is related to the scattering parameters as they contain information of the potential. Consider two waves that are initially in phase.  One wave is sent through an interacting system (or interacts with a target), while the other is sent through free space.  When the first wave emerges from the interacting system (or target), it will no longer be in phase with the second wave.  Thus, we can describe a relation between the potential associated with an interacting system and the phase shift such a system imparts on particles. Due to that, the relations we will find will depend on the potential and the wave function, both known quantities.

And this is precisely what we will find in this section. Afterwards, we will see that if we expand the expression of the 
phase shift for low energies, the coefficients of the expansion are the 
scattering parameters, which are obviously related to the potential. We begin 
with  the radial Schr\"odinger equation for a central 
potential $V(r)$:
\begin{equation}
    u_l''(r)+\left( k^2-U(r)-\frac{l(l+1)}{r^2} \right) u_l(r)=0 \ ,
    \label{schro}
\end{equation}
    with   $k=\sqrt{2\mu E/ \hbar^2}$ and $U(r)=2\mu V(r)/\hbar^2$, $\mu$ being the reduced mass and  $u_l(r)$ the reduced radial wave function.
When there is no potential and the energy is zero, the solution of Eq. (\ref{schro}) is just a combination of power functions:
\begin{equation}
    u_l(r)= N \left[ r^{l+1}+Mr^{-l} \right] \ ,
    \label{eq:energiazero}
\end{equation}
with $N$ and $M$ two integration constants to determine. If the potential is zero but the energy is not, the wave function is a combination of spherical Bessel functions:
\begin{equation}
    u_l(r)=Arj_l(kr)+Brn_l(kr) \ ,
    \label{energianozero}
\end{equation}
with $A$ and $B$ two integration constants to determine.
The boundary condition that we will always apply is that the wave function must 
vanish at $r=0$. This must happen because we are using the reduced wave function, defined as $rR(r)$, where $R(r)$ is the radial part of the true wave function. If we want $R(r)$ to be finite at the origin, $u_l(0)$ must be zero. If  $u_l(0)=0$, then   $B=0$. We recall that the spherical Bessel function $n_l(kr)$ is divergent at the origin, while $j_l(kr)$ is finite. Then $u_l(r)$ reduces 
to
\begin{equation}
    \label{eq:zerosol}
     u_l(r)=Arj_l(kr) \ .
\end{equation}
At large distances, the potential can be neglected, letting us more easily compare the solutions in order to find the phase shift. Eq. 
(\ref{energianozero}) is the wave coming out from an 
interacting system at large distances as it is the general solution, while Eq. (\ref{eq:zerosol}) is the 
 wave coming out from the non-interacting system with the boundary 
condition explained above as in this case the wave function exists in all the space. As we want to compare what happens at large distance, we 
need to know the asymptotic behavior of the Bessel functions for large values of $kr$:
\begin{eqnarray}
    j_l(kr) & \approx & \frac{1}{kr}\sin{\left(kr-\frac{l\pi}{2}\right)}
    \label{besseljn}\\
    n_l(kr) & \approx & \frac{-1}{kr}\cos{\left(kr-\frac{l\pi}{2}\right)} \ .
    \label{besseljn2}
\end{eqnarray}
If we substitute Eqs. (\ref{besseljn}) and (\ref{besseljn2}) into Eq. (\ref{energianozero}), we obtain the asymptotic behavior of $u_l(r)$: 
\begin{equation}
    u_l(r) =\overline{N}\frac{1}{k}\left[ B_1\sin \left( kr-\frac{l\pi}{2}\right)- B_2\cos \left( kr-\frac{l\pi}{2}\right) \right] \ .
        \label{eq:trobarfase1}
\end{equation}
where $A\equiv \overline{N}B_1$ and $B\equiv \overline{N}B_2$, with $\overline{N}$ a constant that we will determine later on.

The phase shift $\delta$ is determined by considering that the large-distance 
solution of the interacting system is the same as the solution with no potential 
at the origin, but with an additional phase. Hence, we include it in equation 
(\ref{eq:zerosol})):
\begin{equation}
    u_l(r)=\overline{A}rj_l(kr+\delta)\approx
    \overline{A}\frac{1}{k}\left[ \sin \left(kr-\frac{l\pi}{2}\right) \cos \delta +\cos\left( kr-\frac{l\pi}{2}\right)  \sin \delta \right] \ .
    \label{eq:trobarfase2}
\end{equation}
In Eq. (\ref{eq:trobarfase2}), we use $\overline{A}$ instead of $A$ to keep the constant $A$ coming from Eq. (\ref{eq:zerosol}) different from the one coming from Eq. (\ref{energianozero}).
If we compare equations (\ref{eq:trobarfase1}) and (\ref{eq:trobarfase2}), we can see that $B_1=\cos{\delta}$, $B_2=-\sin{\delta}$ and $\overline{N}=\overline{A}$. Therefore, the large-distance solution of the interacting system in terms of the phase shift is:
\begin{equation}
    u_l(r)=\overline{N}r\left[ j_l(kr)\cos \delta -n_l(kr)\sin \delta \right]  \ .
    \label{eq:llarguesdistancies0}
\end{equation}
We need to be sure that the constant $\overline{N}$ makes this solution  (\ref{eq:llarguesdistancies0}) compatible with the zero-energy solution  (\ref{eq:energiazero}). We want this because we want our large-distance solution to be also valid at zero energy. To this end, we introduce the $x \to 0$ expansions of the Bessel functions, with $x$ being a generic coordinate:
\begin{eqnarray}
    j_l(x) & \approx & \frac{2^ll!}{(2l+1)!}x^l\bigg[1-\frac{x^2}{2(2l+3)}\bigg]=A_lx^l\bigg[1-\frac{x^2}{2(2l+3)}\bigg] \ ,
    \label{jlsmall}\\
    n_l(x) & \approx & -\frac{(2l)!}{2^ll!}x^{-l-1}\bigg[1+\frac{x^2}{2(2l-1)}\bigg]=-B_lx^{-l-1}\bigg[1+\frac{x^2}{2(2l-1)}\bigg] \ .
    \label{nlsmall}
\end{eqnarray}
Inserting (\ref{jlsmall}) and (\ref{nlsmall}) into Eq. (\ref{eq:llarguesdistancies0}) yields the low-energy limit of the solutions at large distance, we have set $x=kr$:
\begin{equation}
    u_l(r)
    \approx \overline{N}A_lk^l\cos{\delta} \left[ r^{l+1}+\frac{B_l}{A_l}r^{-l}k^{-2l-1}\tan{\delta}\right] \ .
    \label{eq:energiafin}
\end{equation}
The previous limit can be done because, although we are in the range of large $r$, $r$ is supposed to be finite while the energy is tending to zero.
If we compare equations (\ref{eq:energiazero}) and (\ref{eq:energiafin}), we can
relate the integration constants:
\begin{equation}
    \overline{N}=\frac{Nk^{-l}}{A_l\cos{\delta}}\quad ; \quad M=\lim_{k\to0}\frac{B_l}{A_l}k^{-2l-1}\tan{\delta}   \ .
\end{equation}
In the limit $k \to 0$, the constant $M$ is not trivial and  we need to introduce the definition of the cotangent expansion:
\begin{equation}
    k^{2l+1}\cot{\delta}\approx\frac{B_l}{A_l}\frac{1}{a_l^{2l}}\bigg[-\frac{1}{a_l}+\frac{1}{2}r_l^{\text{eff}}k^2\bigg] \ .
    \label{eq:cotangent}
\end{equation}
This expansion (Eq. (\ref{eq:cotangent})) is already known and it can be found elsewhere  \cite{roman,Bishop,HS-2body,newton,genecoef}. As we are in the limit $k \to 0$, if we want Eq. (\ref{eq:cotangent}) to be finite, the cotangent must diverge.
In Eq. (\ref{eq:cotangent}), $a_l$ is the scattering length and 
$r_l^{\text{eff}}$ is the effective range. Right now, the scattering parameters are only the coefficients of the expansion, later on, at the end of Sec. III and in Sec. IV we will give them a physical meaning. Although we have introduced here the expansion of the cotangent without proving it, we show that the tangent, and hence the cotangent, can be expanded in the manner of Eq. (\ref{eq:cotangent}) at the beginning of Sec. \ref{secgen} in Eq. (\ref{prooftan}). However, we point out that Eq. (\ref{eq:cotangent}) is not usually defined as we have done. In some of the references given above, it is written as
\begin{equation}
    k^{2l+1}\cot{\delta}\approx-\frac{1}{a_l^*}+\frac{1}{2}r_l^{\text{eff},*}k^2 \ .
    \label{eq:cotangent2}
\end{equation}
The reason why we do not use Eq. (\ref{eq:cotangent2}) is the fact that $a_l^*$ and $r_l^{\text{eff},*}$ only represent lengths for $l=0$. For $l>0$, these scattering parameters do not have dimensions of length. However, using Eq. (14), $a_l$ and $r_l^{\text{eff}}$ always have dimensions of length. Additionally, we factor out $B_l/A_l$ because we want $a_l=R$ for any $l$ in the case of a hard-sphere potential with diameter $R$ (see Sec. II of Supplementary material II \cite{sup-mat-II}).
As we mentioned right below Eq. (\ref{eq:cotangent}), the 
scattering parameters are the coefficients of the expansion: $a_l$ is the 
coefficient of the zero order term, and $r_l^{\text{eff}}$ is the one 
proportional to $k^2$. If we invert equation (\ref{eq:cotangent}) and we apply 
the limit $k \to 0$, we can find $M$:
\begin{equation}
    M=\frac{B_l}{A_l}\bigg(-\frac{A_l}{B_l}a_l^{2l+1}\bigg)=-a_l^{2l+1}  \ .
\end{equation}
With that, the solution at zero energy ($k=0$) becomes
\begin{equation}
    u_l(r,k=0)=N\Big[r^{l+1}-a_l^{2l+1}r^{-l}\Big]
 \ ,
 \end{equation}
and at finite energy,
\begin{equation}
    u_l(r)=\frac{Nk^{-l}}{A_l}r\Big[j_l(kr)-n_l(kr)\tan{\delta}\Big] \ .
    \label{eq.funconak}
\end{equation}
The next step is to relate the phase shift $\delta$ to the potential $U(r)$. In order to do so, we shall combine the true differential equation with the one that the regular Bessel function satisfies:
\begin{eqnarray}
    \bigg[(rj_l(kr))''+\bigg(k^2-\frac{l(l+1)}{r^2}\bigg)rj_l(kr)=0\bigg] & \times & u_l(r) \ , \\
    \bigg[u_l''(r)+\bigg(k^2-U(r)-\frac{l(l+1)}{r^2}\bigg)u_l(r)=0\bigg] & \times & (rj_l(kr)) \ .
\end{eqnarray}
By subtracting the above equations one gets
\begin{equation}
\label{eq.derischro}
    \frac{d}{dr}\bigg((rj_l(kr))'u_l(r)-rj_l(kr)u_l'(r)\bigg)=-U(r)rj_l(kr)u_l(r) \ .
\end{equation}
We can integrate now the entire equation between zero and infinity for  
regular (physical) potentials:
\begin{equation}
    \bigg[(rj_l(kr))'u_l(r)-rj_l(kr)u_l'(r)\bigg]_{0}^{\infty}=-\int_0^{\infty}U(r)rj_l(kr)u_l(r)dr \ .
\end{equation}
When evaluating the left-hand side, the value at $r=0$ is zero, so we need only the value at $r\rightarrow\infty$:
\begin{equation}
    \bigg[(rj_l(kr))'u_l(r)-rj_l(kr)u_l'(r)\bigg]_{\infty}=-\int_0^{\infty}U(r)rj_l(kr)u_l(r)dr \ .
\end{equation}
Using the asymptotic approximation for the Bessel functions (\ref{besseljn},\ref{besseljn2}), 
the wave function and its derivative at infinity are:
\begin{eqnarray}
    \lim_{r\rightarrow\infty}u_l(r) & = & \frac{Nk^{-l}}{A_l}\frac{1}{k}\bigg[\sin{\bigg(kr-\frac{l\pi}{2}\bigg)}+\cos{\bigg(kr-\frac{l\pi}{2}\bigg)}\tan{\delta}\bigg]_{\infty} \ , \\ 
    \lim_{r\rightarrow\infty}u_l'(r) & = & \frac{Nk^{-l}}{A_l}\bigg[\cos{\bigg(kr-\frac{l\pi}{2}\bigg)}-\sin{\bigg(kr-\frac{l\pi}{2}\bigg)}\tan{\delta}\bigg]_{\infty}  \ .
\end{eqnarray}
After some algebraic manipulations, we find:
\begin{equation}
    \bigg[(rj_l(kr))'u_l(r)-rj_l(kr)u_l'(r)\bigg]_{\infty}=\frac{Nk^{-l}}{A_l}\frac{1}{k}\tan{\delta} \ ,
\end{equation}
and the tangent of the phase shift is:
\begin{equation}
    \tan{\delta}=-A_lk^{(l+1)} \int_0^{\infty}U(r)rj_l(kr)\frac{u_l(r)}{N}dr  \ 
.
    \label{eq:tangent}
\end{equation}
With the expression (\ref{eq:tangent}) we can know the phase shift, but we need to relate it with the cotangent expansion (Eq. (\ref{eq:cotangent})) because our goal is to find the scattering parameters. Therefore, we calculate the cotangent from the expression (\ref{eq:tangent}) by considering an expansion. First, we consider that the tangent can be written as 
\begin{equation}
    \tan{\delta}\approx-\frac{A_l}{B_l}k^{2l+1}\big[c_1+\frac{1}{2}c_2k^2\big]
    \label{eq.supostan} \ ,
\end{equation}
Eq. (\ref{eq.supostan}) is valid for low energies (low $k$), and thus,
\begin{equation}
    \cot{\delta}=-\frac{B_l}{A_l}\frac{1}{k^{2l+1}}\frac{1}{c_1}\frac{1}{1+\cfrac{c_2}{2c_1}k^2}\approx -\frac{B_l}{A_l}\frac{1}{k^{2l+1}}\frac{1}{c_1}\bigg(1-\frac{1}{2}\frac{c_2}{c_1}k^2\bigg) \ .
\end{equation}
Finally, by comparing with Eq. (\ref{eq:cotangent}), we can relate the coefficients $c_1$ and $c_2$ in expansion (\ref{eq.supostan}) to the scattering parameters,
\begin{equation}
    c_1 = a_l^{2l+1}
    \quad \quad c_2=a_l^{2l+2} r_l^{\text{eff}} \ .
    \label{tancoe}
\end{equation}
Up to this point, we have related the tangent of the phase shift to 
the potential and the wave function through an integral (\ref{eq:tangent}). The 
scattering parameters appear as coefficients of the low-energy expansion of 
the tangent of the phase shift.

\section{Scattering parameters}
\label{secgen}
In this section, we will find the expressions for the scattering parameters in terms of the potential and the wave function.
In Eq. (\ref{eq.supostan}), we split the expression of the tangent into two terms, one proportional to $k^{2l+1}$ and another proportional to $k^{2l+3}$, whose coefficients have been found at the end of the previous section. Now, we break Eq. (\ref{eq:tangent}) in two parts in order to compare both expressions and find the coefficients, hence finding the scattering parameters.

Everything done in this section is under the assumption of a regular potential, that is, a potential with no divergences (except at the origin) that decays faster than $1/r^n$ with $n>2l+3$ for each partial wave at infinity (Wigner threshold).
We will break the integral:
\begin{equation}
    \tan{\delta}=-A_lk^{l+1}\int_0^{\infty}U(r)rj_l(kr)\frac{u_l(r)}{N}dr \ ,
\end{equation}
in two parts in order to obtain terms proportional to $k^{2l+1}$ and $k^{2l+3}$.
To this end,  we  expand the Bessel function (Eq. (\ref{jlsmall})) and the wave function:
\begin{equation}
    u_l(r)=N\Big(u_l^{(0)}(r)+k^2 u_l^{(2)} (r)+ \ldots \Big) \ ,
      \label{uapproach}
    \end{equation}
around $k=0$. The super-indices in Eq.(\ref{uapproach}) indicate the order 
of the expansion. The series (\ref{uapproach}) has only even terms. The odd 
terms are zero because the radial Schr\"odinger equation depends on $k$ as $k^2$, 
causing $u_l(r,k) = u_l(r,-k)$. With 
this, one obtains
\begin{equation}
    \tan{\delta}
    \approx -\frac{A_l}{B_l}k^{2l+1}\bigg[A_lB_l\int_0^{\infty}U(r)r^{l+1}u_l^{(0)}(r) dr+\frac{1}{2}(2A_lB_l)k^2\int_0^{\infty}U(r)r^{l+1}\bigg(u_l^{(2)}(r)-\frac{r^2 u_l^{(0)}(r)}{2(2l+3)}\bigg)dr\bigg]  \ .
    \label{prooftan}
\end{equation}
Comparing the latter equation with Eq. (\ref{eq.supostan}), one can 
identify the coefficients $c_1$ and $c_2$ from Eq. (\ref{tancoe}): 
\begin{eqnarray}
    c_1 & = & \frac{1}{2l+1}\int_0^{\infty}U(r)r^{l+1}u_l^{(0)}(r) dr \ , \\
    c_2 & = & 
\frac{2}{2l+1}\int_0^{\infty}U(r)r^{l+1}\bigg(u_l^{(2)}(r)-\frac{r^2 
u_l^{(0)}(r)}{2(2l+3)}\bigg)dr    \ , 
    \label{eq:constant2}
\end{eqnarray}
where we have used $A_l B_l =1/(2l +1)$, which can be derived from Eqs. (\ref{jlsmall}) and (\ref{nlsmall}). 
In this way, we already get $a_l$ because $c_1=a_l^{2l+1}$. To continue with 
$c_2$ and thus to find $r_l^{\text{eff}}$, we need to know what equations 
$u_l^{(0)}(r)$ and $u_l^{(2)}(r)$ satisfy. To do so we plug the expansion of the 
wave function (\ref{uapproach}) into the Schr\"odinger equation:
\begin{equation}     
u_l^{(0)''}(r)+k^2u_l^{(2) 
''}(r)+\bigg(k^2-U(r)-\frac{l(l+1)}{r^2}\bigg)\Big(u_l^{(0)} 
(r)+k^2u_l^{(2)}(r)\Big)=0
\end{equation}
At order $k^0$ the differential equation is:
\begin{equation}
    u_l^{(0)''}(r)-\bigg[\frac{l(l+1)}{r^2}+U(r)\bigg]u_l^{(0)}(r)=0
    \label{u0order}
\end{equation}
and thus $u_l^{(0)}(r)$ is nothing else than the wave function at zero energy. 
Here, we can understand the Wigner threshold: if we want $c_1$ to exist, the 
potential must decay faster than $1/r^n$ with $n>2l+3$ for each partial wave. 
We recall that $u_l^{(0)}(r)$ at infinity goes as $r^{l+1}-a_l^{2l+1}r^{-l}$. At 
order $k^2$, one gets:
\begin{equation}
    u_l^{(2)''}(r)-\bigg[\frac{l(l+1)}{r^2}+U(r)\bigg]u_l^{(2)}(r)=-u_l^{(0)}(r) \ .
\label{u2order}
    \end{equation}
It looks similar to the one for $u_l^{(0)}(r) $ (\ref{u0order}) but now there is a source term, which turns out to be $u_l^{(0)}(r)$ itself. Combining both equations:
\begin{eqnarray}
    \bigg[u_l^{(0)''}(r)-\bigg(\frac{l(l+1)}{r^2}+U(r)\bigg) u_l^{(0)}(r) =0\bigg] 
& \times & u_l^{(2)}(r)\\
    \bigg[u_l^{(2)''}(r)-\bigg(\frac{l(l+1)}{r^2}+U(r)\bigg)u_l^{(2)}(r)=-u_l^{(0)}
(r)\bigg ] & \times & u_l^{(0)}(r)
\end{eqnarray}
and integrating, one obtains:
\begin{equation}
    \left[ u_l^{(0)'}(r) u_l^{(2)}(r) - u_l^{(2)'}(r) u_l^{(0)}(r) 
\right]_0^r=\int_0^{r}{u_l^{(0)}}^2(r)dr  \ .
\end{equation}
The functions $u_l^{(0)}(r)$ and $u_l^{(2)}(r)$ are zero at the origin for any energy because $u_l(r)$ is zero at the origin due to boundary conditions. As $u_l^{(0)}(r)$ and $u_l^{(2)}(r)$ are the terms in the expansion of $u_l(r)$ (Eq. (\ref{uapproach})), if $u_l(r)$ is zero at the origin for any energy, then each term of the expansion must be also zero at the origin.
Therefore:
\begin{equation}
    u_l^{(2)'}(r)=\frac{u_l^{(0)'}(r)}{u_l^{(0)}(r)} u_l^{(2)}(r)-\frac{1}{u_l^{(0)}(r)} 
\int_0^{r} {u_l^{(0)}}^2(r)dr \ .
    \label{eq:devu2}
\end{equation}
Using the differential equations for $u_l^{(0)}(r)$ and $u_l^{(2)}(r)$  we can rewrite the constant $c_2$ from Eq. (\ref{eq:constant2}) in the following way (see Sec. I of Supplementary material I \cite{sup-mat-I} for the explicit derivation):
\begin{equation}
    \frac{(2l+1)c_2}{2}=r^{l+1}u_l^{(2)'}(r)\Big|_0^{\infty}-(l+1)r^lu_l^{(2)}(r) 
\Big|_0^{\infty}+\frac{1}{2(2l+3)}\bigg\{(l+3)r^{l+2}u_l^{(0)}(r)\Big|_0^{\infty}
-r^{l+3}u_l^{(0)'}(r)\Big|_0^{\infty}\bigg\} \ .
    \label{eq:app1}
\end{equation}
In Eq. (\ref{eq:app1}), everything is zero in the inferior limit $r=0$. Substituting $u_l^{(2)'}(r)$ by Eq. (\ref{eq:devu2}), we get:
\begin{equation}    
\frac{(2l+1)c_2}{2}=\lim_{r\rightarrow\infty}\bigg\{\frac{(l+3)r^{l+2}u_l^{(0)}}{2(2l+3)}-\frac{r^{l+3}u_l^{(0)'}}{
2(2l+3)}-\frac{u_l^{(2)}}{u_l^{(0)}}[(l+1)r^l 
u_l^{(0)}-r^{l+1}u_l^{(0)'}]-\frac{r^{l+1}}{u_l^{(0)}}\int_0^{r}{u_l^{(0)}}^2(r)dr\bigg\} \ .
    \label{eq:constant2_2}
\end{equation}
In order to continue, we need to know $u_l^{(0)}(r)$ and $u_l^{(2)}(r)$ in the limit $r \to \infty$. 
To this end, we combine Eqs. (\ref{eq.funconak}), (\ref{eq.supostan}), (\ref{jlsmall}) and (\ref{nlsmall}) to write:
\begin{equation}
    u_l(r) =N\bigg\{r^{l+1}-a_l^{2l+1}r^{-l}+k^2\bigg[-\frac{r^{l+3}}{2(2l+3)}-a_l^{2l+1}\frac{r^{2-l}}{2(2l-1)}-\frac{1}{2}a_l^{2l+2}r_l^{\text{eff}}r^{-l}\bigg]\bigg\} \ .
    \label{k0k2}
    \end{equation}
In Eq. (\ref{k0k2}), the term proportional to $k^0$ is just $u_l^{(0)}(r)$ and the one proportional to $k^2$ is $u_l^{(2)}(r)$:
\begin{eqnarray}
    u_l^{(0)}(r) & = & r^{l+1}-a_l^{2l+1}r^{-l} \\
    u_l^{(2)}(r) & = & -\frac{r^{l+3}}{2(2l+3)}-a_l^{2l+1}\frac{r^{2-l}}{2(2l-1)}-\frac{1}{2}a_l^{2l+2}r_l^{\text{eff}}r^{-l} \ .
\end{eqnarray}
The next step is the substitution of $u_l^{(0)}$, $u_l^{(0)'}$ and $u_l^{(2)}$ into 
the expression of $c_2$ (\ref{eq:constant2_2}). After some manipulations (see Sec. II of Supplementary material I \cite{sup-mat-I} for further details), we obtain:
\begin{equation}
    c_2=\frac{2}{2l+1}\lim_{r\rightarrow\infty}\bigg\{\frac{r^{2l+3}}{2l+3}-a_l^{2l+1}r^2+\frac{a_l^{4l+2}r^{1-2l}}{1-2l}-\int_0^{r}{u_l^{(0)}}^2(r)dr \bigg\}   \ .
\end{equation}
Finally, considering the relation between $c_2$ and $r_l^{\text{eff}}$, 
we obtain the effective range:
\begin{equation}
    r_l^{\text{eff}}=\frac{c_2}{a_l^{2l+2}}=\frac{2}{(2l+1)a_l^{2l+2}}
\lim_{r\rightarrow\infty}\bigg\{\frac{r^{2l+3}}{2l+3}-a_l^{2l+1}r^2+\frac{a_l^{ 
4l+2}r^{1-2l}}{1-2l}-\int_0^{r}{u_l^{(0)}}^2(r)dr\bigg\} \ .
\label{eq.apprange1}
\end{equation}
Applying the limit $r \to \infty$ in the above expression yields two different expressions depending on whether $l=0$ or $l>0$. The reason for this difference is the term containing a distance to the power $(1-2l)$. This term goes to zero at infinity if $l >0$; however if $l=0$, this term diverges. The expressions after this split are:
\begin{equation}
   r_0^{\text{eff}}=\frac{2}{a_0^2}\lim_{r\rightarrow\infty}\bigg[\frac{r^3}{3}
-a_0r^2+a_0^2r-\int_0^{r}{u_0^{(0)}}^2(r)dr\bigg] \ ,
\end{equation}
\begin{equation}
 r_{l>0}^{\text{eff}}=\frac{2}{(2l+1)a_l^{2l+2}}\lim_{r\rightarrow\infty}\bigg[ 
\frac { r^ {2l+3}}{2l+3}-a_l^{2l+1}r^2-\int_0^{r}{u_l^{(0)}}^2(r)dr\bigg] \ .
\end{equation}
All the power terms give zero at $r=0$. That means we can write the full expressions in integral form, containing the zero as the inferior limit, and take the limit of $r \to \infty$. Finally, the resulting expressions for the effective range for any $l$ are:
\begin{equation}
\boxed{
 r_0^{\text{eff}}=\frac{2}{a_0^2}\int_0^{\infty}\Big[(r-a_0)^2-{u_0^{(0)}}
^2(r)\Big ] dr  }  
    \label{eferangel0}
\end{equation}
\begin{equation}
\boxed{
   r_{l>0}^{\text{eff}}=\frac{2}{(2l+1)a_l^{2l+2}}\int_0^{\infty}\Big[r^{2l+2}
-2a_l^ { 2l+1 }r-{u_l^{(0)}}^2(r)\Big]dr
    }
    \label{eferangelm0}
\end{equation}
For any $l$, the integrand vanishes in the limit $r\rightarrow\infty$ since in 
this limit the wave function squared matches the power functions. If the wave 
function were to have this behavior everywhere, not only at infinity, the 
effective range would be zero. As in general this is not the case, the effective 
range takes into account the separation between the true solution and the 
dominant terms of the large distance scattering solution.

The above formulas are equivalent to the ones found by L. B. Madsen in Ref. 
\cite{lars}:
\begin{equation}
r_l^{\text{eff},*}=\lim_{k\rightarrow0}2\bigg(\int_0^Rdr[v_{l,0}(r)v_{l,k}(r)-u_{l
, 0 } 
(r)u_{l,k}(r)]-\frac{1}{a_l^*}\frac{R^2}{(2l+1)}-\frac{[(2l-1)!!]^2}{(2l-1)R^{2l-1}} 
\bigg) \ .
    \label{eq.rangeffmadsen}
\end{equation}
In Ref. \cite{lars}, $u_{l,k}(r)$ is the solution to the radial Schr\"odinger equation for any energy (that is our $u_l(r)$); $u_{l,0}(r)$ is the same but at zero energy (that is our $u_l^{(0)}(r)$); $v_{l,k}(r)$ is the solution to the radial Schr\"odinger equation for any energy when the potential is zero and that is finite at the origin (it is the function that we have at Eq. (\ref{eq:zerosol})); and finally, $v_{l,0}(r)$ is the limit of zero energy of the previous function which is a constant times $r^{l+1}$. All the functions are taken to be zero at the origin. 
Differences in our approach and that of Ref. \cite{lars} to normalizing $u_{l,k}(r)$ and $v_{l,k}(r)$ are described in Sec. III of the Supplemental material I \cite{sup-mat-I}.

Our expressions for the effective range are simpler to use than the ones from Ref. \cite{lars} (our Eq. (\ref{eq.rangeffmadsen})), since they only include integrals that converge, whereas Eq. (\ref{eq.rangeffmadsen}) additionally requires setting the limits $k \to 0$ and $R \to \infty$. Apart from that, the effective ranges are defined in two slightly different ways (see Sec. III of Sup. material I \cite{sup-mat-I}).

For completeness, the formula for the $l$-wave scattering length $a_l$ is:
\begin{equation}
\label{equacioal}
\boxed{
    a_l^{2l+1}=\frac{1}{2l+1}\int_0^{\infty}U(r)r^{l+1} u_l^{(0)}(r) dr \ .
    }
\end{equation}
Notice that that Eqs. (\ref{eferangel0},\ref{eferangelm0},\ref{equacioal}) need the wave function $u_l^{(0)}$, which is the zero-energy solution of the radial Schr\"odinger equation with the boundary condition $u_l^{(0)}(r=0)=0$ (the wave function must vanish at the origin) and properly normalized to behave as $r^{l+1}-a_l^{2l+1}r^{-l}$ when $r\rightarrow\infty$.

\section{Resonance scattering}

In this section, we discuss the 
meaning of positive and negatives values of the scattering length and when 
divergences appear. 

\begin{figure}
    \centering
    \begin{subfigure}[b]{0.49\textwidth}
        \centering
        \includegraphics[width=\textwidth]{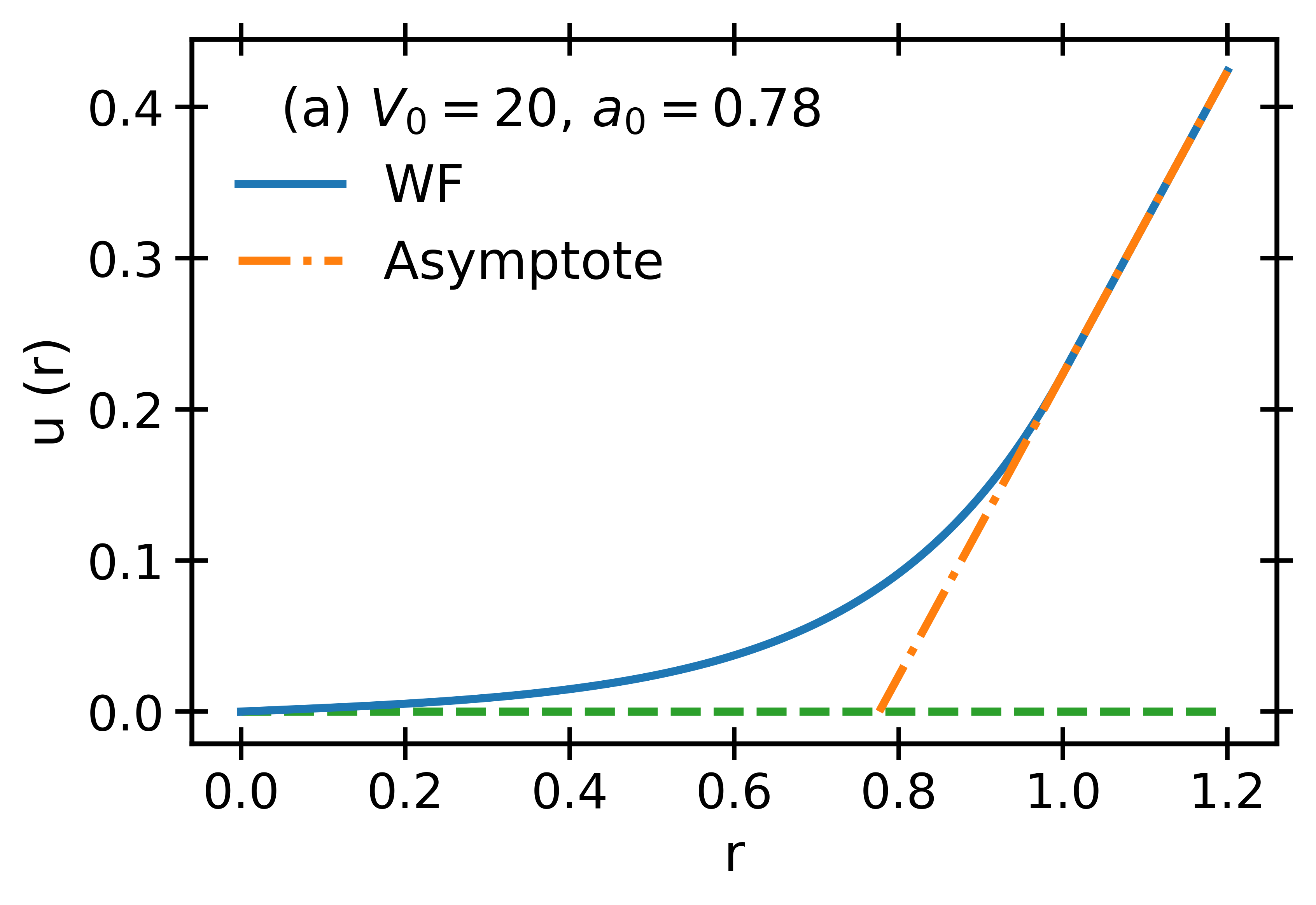}
        \captionlistentry{}
        \label{figa}
    \end{subfigure}
    \begin{subfigure}[b]{0.49\textwidth}
        \centering
        \includegraphics[width=\textwidth]{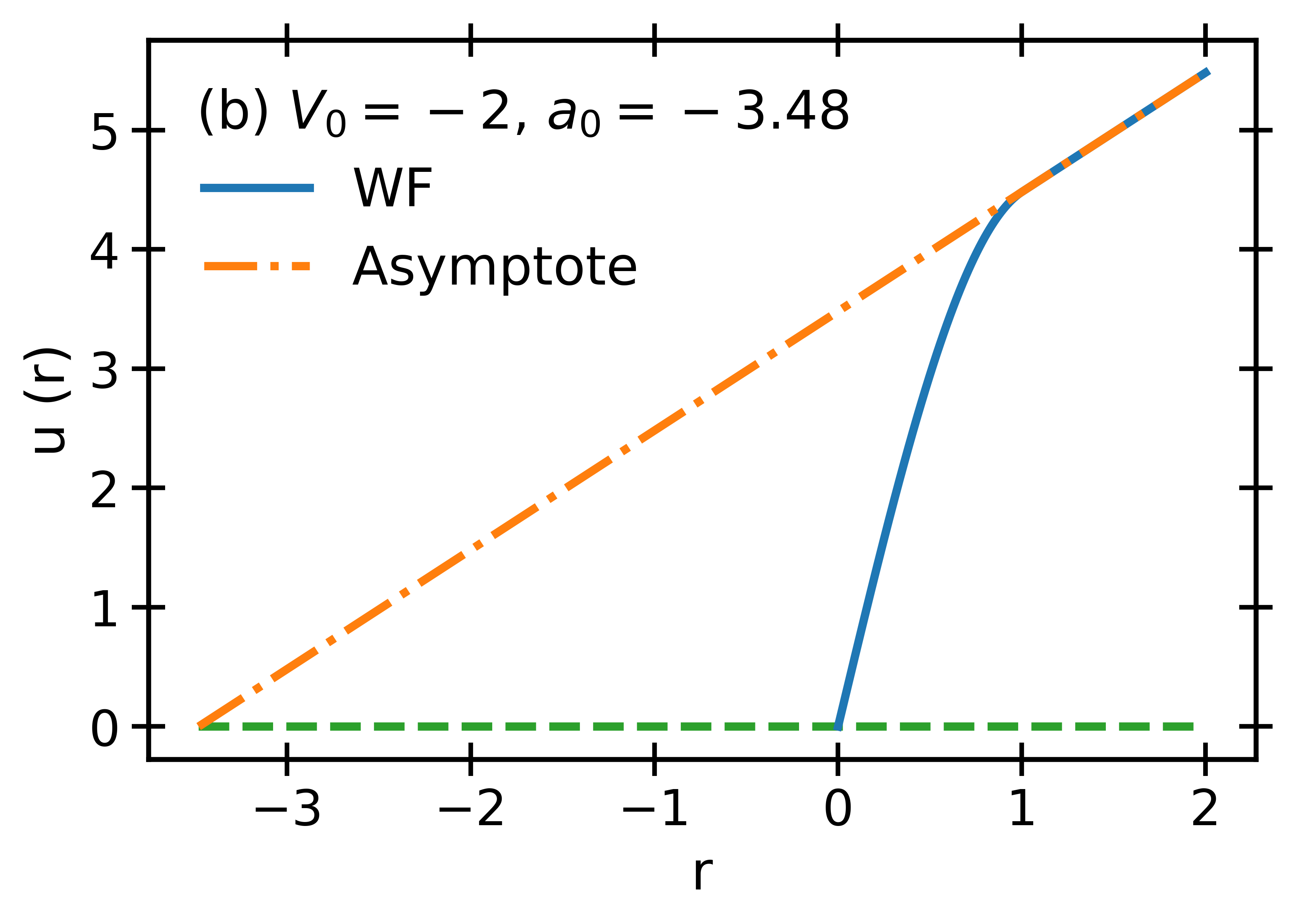}
        \captionlistentry{}
        \label{figb}
    \end{subfigure}
    \\
    \begin{subfigure}[b]{0.49\textwidth}
        \centering
        \includegraphics[width=\textwidth]{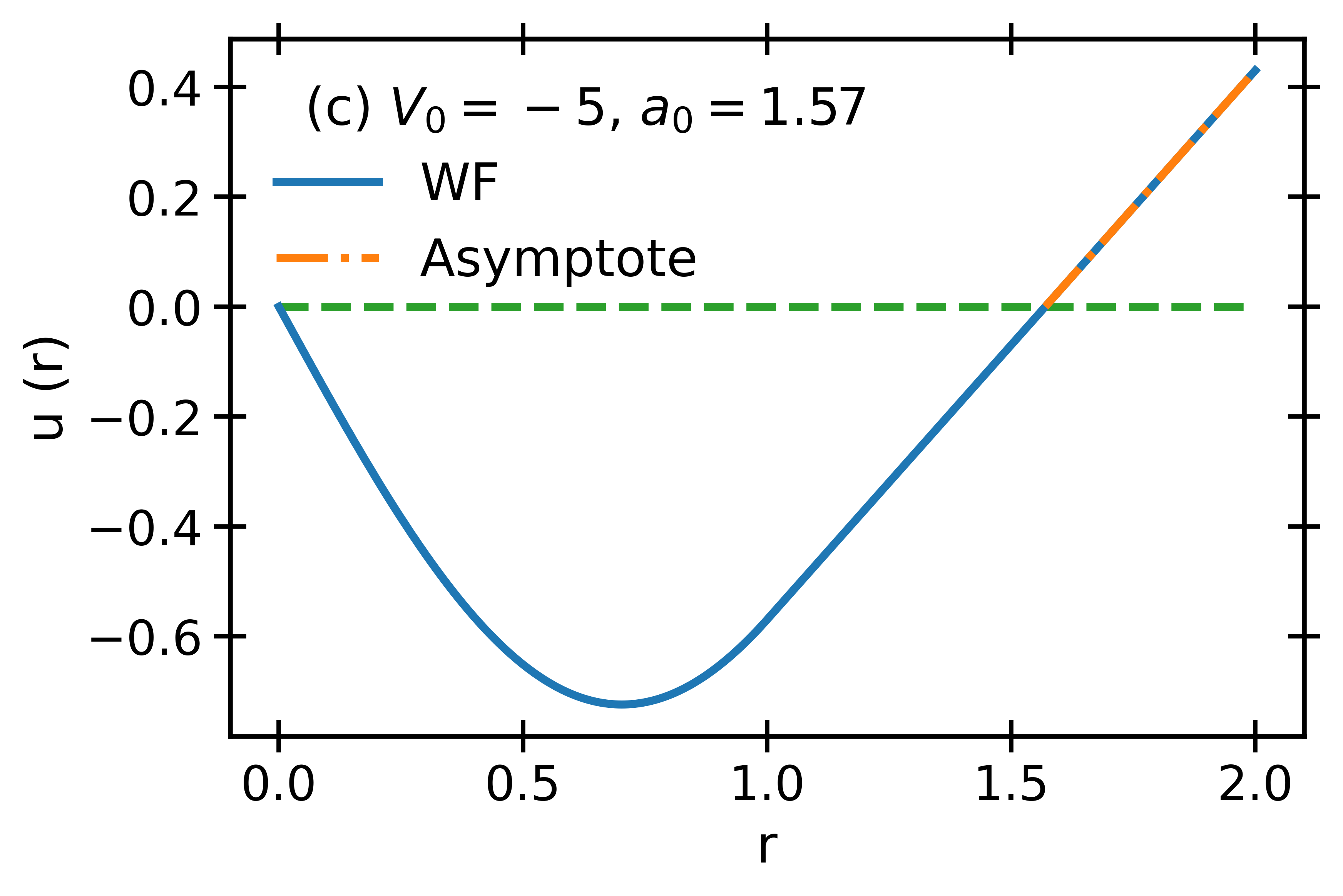}
        \captionlistentry{}
        \label{figc}
    \end{subfigure}
    \begin{subfigure}[b]{0.49\textwidth}
        \centering
        \includegraphics[width=\textwidth]{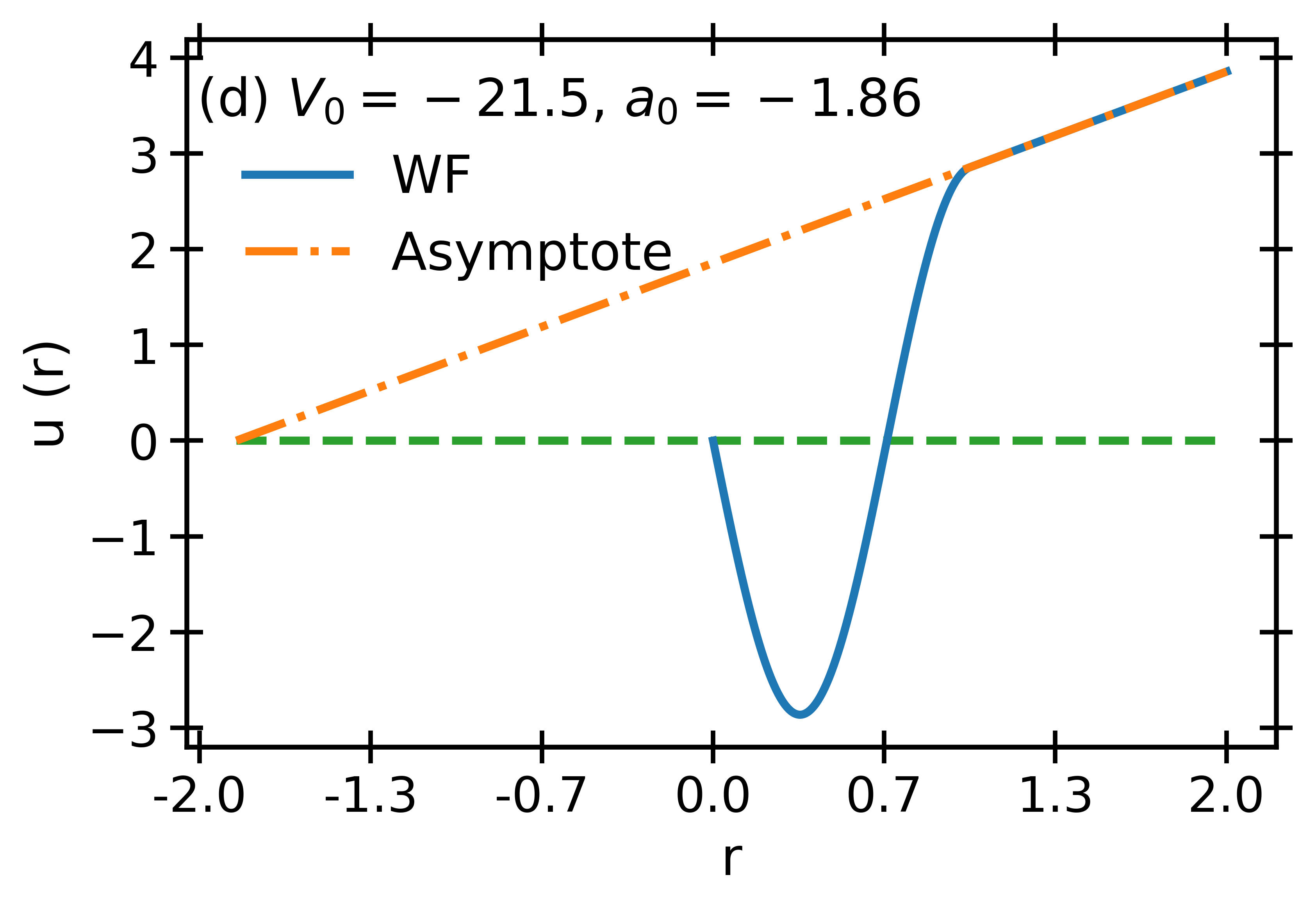}
        \captionlistentry{}
        \label{figd}
    \end{subfigure}
    \caption{\footnotesize{The wave functions (solid blue lines) are shown for four different potentials, along with linear approximations to their behavior at large $r$ (orange dotted lines).  The point at which the linear approximation crosses zero yields the value of the scattering length, $a_0$.  The potentials are all zero for $r > 1$ and take the values $V_0$=20, -2, -5, and -21.5 for $r < 1$.}}
\label{graf_WF}
\end{figure}

For any $l$, the scattering length is the point where  
the large-distance behavior 
of the wave function extrapolates to a value of zero, as seen in Fig. \ref{graf_WF}. This fact comes from the large distance solution $u_l(r)=r^{l+1}-a_l^{2l+1}r^{-l}$. If $r=a_l$, $u_l(r)$ is zero. This picture 
allows us to have an idea about the scattering length without looking at the 
explicit formulas. For example, if the potential is repulsive, then at distances close 
to the origin  the wave function will be 
small, and as we move away from it, the wave function will increase. We will 
have  a concave up function and  thus  the 
intersection of the large-distance solution with the horizontal axis will be 
positive. This can also be verified looking at Eq. (\ref{equacioal}): if 
all the terms inside the integral are positive, the integral, and hence $a_l$, 
will be positive.

The case of attractive potentials is richer 
than that of repulsive potentials since the strength of the attractive potential must be considered. Weakly attractive potentials exhibit the opposite behavior of repulsive potentials. In the weakly attractive case, the 
tendency to be near the origin is larger, making the wave function to be 
concave down with a negative scattering length. If we increase the 
strength of the potential, the concavity will decrease and the scattering length 
will become more negative. Once we reach a critical value of the potential, the 
intersection distance will be at negative infinity. This point is particularly interesting, and we will comment on it later on (Feshbach resonance). After 
crossing this point, the scattering length becomes divergent, but at positive 
infinity. From this situation, the evolution will be the same as the one we 
have explained: with increasing potential, the scattering length will start to 
diminish, will become zero, and then will start approaching negative infinity. 
After crossing the divergence, the cycle will be repeated. This process is 
illustrated in Fig. (\ref{graf_WF}), where we plot four cases using a 
soft-sphere potential for (\ref{figa}) and a spherical-well 
potential for (\ref{figb}), (\ref{figc}) and (\ref{figd}). The first case (\ref{figa}) corresponds to a repulsive 
potential and the intersection with the $r$ axis is positive. In (\ref{figb}), (\ref{figc}) and (\ref{figd}), we plot 
the wave function for three attractive potentials of increasing strength in 
order to show the regimes explained above. In (\ref{figb}), the potential is 
weak and we have not reached the critical value. In (\ref{figc}), we have 
crossed the Feshbach resonance. In (\ref{figd}), we show a case which is 
before the second Feshbach resonance.

We want to analyze the critical point, that is, the situation in which the 
scattering length diverges, going from negative to positive infinity. The Schr\"odinger equation always has two kind of solutions: scattering states and bound states. Scattering states can have any positive energy 
and at infinity behave like a non-normalizable wave. Bound 
states form a discrete set of solutions that have negative energy, vanish at infinity, and are normalizable. Upon increasing the strength of an 
attractive potential, as bound states are discrete, there will be certain 
values in which a scattering state with very low energy will become a bound 
state with negative energy close to zero. As bound states vanish at infinity, 
they cannot be described with the large-distance solution 
($u_l(r)=r^{l+1}-a_l^{2l+1}r^{-l}$) that we have used to this point. The large-distance solution of a bound state is $Nr^{-l}$, with $N=0$ if $l=0$. In order to be able to normalize the mode, the term $a_l^{2l+1}r^{-l}$ should dominate $r^{l+1}$. This 
anomaly causes the scattering length to diverge. In fact, in this very 
limit (called the unitary limit) the scattering length is undefined because it can 
be $\pm\infty$. This process in which a scattering state transitions to a bound state as the scattering length diverges from negative to positive infinity due to an increasingly attractive potential is called a Feshbach resonance~\cite{chin}.

This divergent behavior around certain values of the potential can be modelled 
if we Taylor-expand the inverse of the scattering length around these values:
\begin{equation}    
\frac{1}{a_l}(V_0)=\frac{1}{a_l}\bigg|_{V_c}+\bigg(\frac{1}{a_l}\bigg)'\bigg|_{
V_c}(V_0-V_c)+\frac{1}{2}\bigg(\frac{1}{a_l}\bigg)''\bigg|_{V_c}(V_0-V_c)^2+ 
\ldots
\end{equation}
Requiring that the inverse of the scattering length at these values is zero, and 
ignoring orders higher than 1, we obtain:
\begin{equation}
    a_l=\frac{C}{V_0-V_c}\quad\text{where}\quad C=\bigg[\bigg(\frac{1}{a_l}\bigg)'\bigg|_{V_c}\bigg]^{-1}
\end{equation}

Feshbach resonances appear when the energy of a scattering state matches the energy of a 
bound state. Although in the procedure explained above we have crossed the 
resonance by means of increasing the strength of the potential, we can leave the 
potential unchanged and make use of external sources such as magnetic fields to vary the effective interaction and cross the resonance either-way. 
Magnetically tunable Feshbach resonances are very important in ultracold 
quantum gases. Their use  provides means of varying the scattering length almost 
at will. \cite{pethick_1} For instance, one can go from a Fermi
gas with pairing (BCS) with $a<0$ to a Bose-Einstein condensate (BEC) with 
$a>0$ by crossing a Feshbach resonance. \cite{pethick_2}

As we have found a compact formula for the effective range, it is interesting 
to explore the behavior of the effective range whenever $a_l$ 
is zero or $\pm\infty$. The case $a_l=0$ is only interesting from a mathematical 
point of view. Physically, it is not relevant because 
what matters is the product of $r_l^{\text{eff}}$ times $a_l^{2l+2}$. This 
product is the constant $c_2$ introduced in Sec. \ref{secII}, which is one of 
the coefficients of the expansion of the tangent of the phase shift. On the 
other hand, the case $a_l=\pm\infty$ is physically important, since this corresponds to the behavior around the 
Feshbach resonance. To this end, we write a compact expression of 
$r_l^{\text{eff}}$:
\begin{equation}
r_l^{\text{eff}}=\frac{2}{(2l+1)a_l^{2l+2}}\int_0^{\infty}\Big[r^{2l+2}-2a_l^{
2l+1 } r+a_l^{
4l+2}r^{-2l}\delta_{l,0}-{u_l^{(0)}}^2(r)\Big]dr \ .
\end{equation}
First, we analyze its behavior when $a_l \to 0$. By looking at the dominant 
terms in this limit, one gets:
\begin{equation}
r_l^{\text{eff}}\rightarrow\frac{2}{(2l+1)a_l^{2l+2}}\int_0^{\infty}\Big[r^{2l+2
} - { u_l^{(0)}}
^2(r)\Big]dr\rightarrow\pm\infty \ .
    \label{rldiverg1}
\end{equation}
The integral in Eq. (\ref{rldiverg1}) is always finite unless $a_l$ 
diverges. For the case where $a_l$ does not diverge, the effective range can only diverge if the fraction preceding the integral diverges, which would necessarily be due to $a_l^{2l+2}$. However, since the product of $a_l^{2l+2}$ and $r_l^{\text{eff}}$ is finite, 
\begin{equation}
r_l^{\text{eff}}a_l^{2l+2}\rightarrow\frac{2}{(2l+1)}\int_0^{\infty}\Big[r^{2l+2
} - { u_l^{(0)}}
^2(r)\Big]dr \ ,
\label{rldiverg2}
\end{equation}
this limit (Eq. (\ref{rldiverg1})) is not physically relevant. We recall that the product $r_l^{\text{eff}}a_l^{2l+2}$ is what appears in the expansion of the tangent of delta, see Eqs. (\ref{eq.supostan}) and (\ref{tancoe}).

When considering the case in which $a_l \to \pm \infty$, it is useful to split the wave function into two terms (the derivation of which is found in Sec. IV of the Supplementary material I \cite{sup-mat-I}):
\begin{equation}
    u_l^{(0)}(r)=\frac{lu_l^{(0)}(r)+r{u'}_l^{(0)}(r)}{1+2l}-\frac{r{u'}_l^{(0)}(r)-(l+1)u_l^{(0)}(r)}{1+2l} \ .
\end{equation}
The first term goes to positive infinity as $r^{l+1}$ while the second term goes to positive infinity as $a_l^{2l+1}r^{-l}$.
We recall that at infinity $u_l^{(0)}(r)$ is $r^{l+1}-a_l^{2l+1}r^{-l}$. As long as $a_l$ is non-zero, we can redefine the two parts as two functions:
\begin{equation}
    \frac{lu_l^{(0)}(r)+r{u'}_l^{(0)}(r)}{1+2l}\equiv f_l(r)\quad;\quad \frac{r{u'}_l^{(0)}(r)-(l+1)u_l^{(0)}(r)}{1+2l}\equiv a_l^{2l+1}h_l(r) \ .
\end{equation}
Both $f_l(r)$ and $h_l(r)$ at $r=0$ are zero. And at $r \to \infty$, they are 
$r^{l+1}$ and $r^{-l}$ respectively. The differential equations that $f_l(r)$ 
and $h_l(r)$ satisfy can be found in Sec. IV of the Supplementary material I \cite{sup-mat-I}. With this, the wave function can be rewritten as 
$u_l^{(0)}(r)=f_l(r)-a_l^{2l+1}h_l(r)$. If we apply this transformation to the 
effective range, we obtain:
\begin{equation}
r_l^{\text{eff}}=\frac{2}{(2l+1)a_l^{2l+2}}\bigg\{\int_0^{\infty}\Big[r^{2l+2}-f_l^2(r)\Big]dr-2a_l^{2l+1}\int_0^{\infty}\Big[r-f_l(r)h_l(r)\Big]+a_l^{
4l+2}\int_0^{\infty}\Big[r^{-2l}\delta_{l,0}-h_l^2(r)\Big]dr\bigg\} \ .
\end{equation}
The above integrals are finite. This can be seen by noting that $f_l^2(0)$, 
$f_l(0)h_l(0)$, and $h_l^2(0)$ are zero, and that 
$\lim_{r\rightarrow\infty}\Big[r^{2l+2}-f_l^2(r)\Big]=0$, 
$\lim_{r\rightarrow\infty}\Big[r-f_l(r)h_l(r)\Big]=0$, and 
$\lim_{r\rightarrow\infty}\Big[r^{-2l}\delta_{l,0}-h_l^2(r)\Big]=0$.

We distinguish two cases, $l=0$ and $l>0$. In the first case, when $a_0 \to 
\pm\infty$:
\begin{equation}
     r_0^{\text{eff}}\rightarrow 
2\int_0^{\infty}\Big[1-h_l^2(r)\Big]dr\rightarrow\text{finite value} \ .
\end{equation}
Instead, for $l>0$ and  $a_l \to \pm\infty$:
\begin{equation}
r_l^{\text{eff}}\rightarrow-\frac{2}{(2l+1)}a_l^{2l}\int_0^{\infty}h_l^2(r)dr\rightarrow-\infty
\ .
\end{equation}
As a summary, whenever $a_l$ is zero, $r_l^{\text{eff}}$ diverges. For a 
diverging scattering 
length, we have two situations: for $l=0$, the effective range is finite, while 
for $l>0$, it diverges to $-\infty$.

\section{Examples}
In this section, we apply the scattering formalism to obtain the scattering length and effective ranges of model potentials that can be analytically integrated. We just present the final formulas, the development can be found in the Supplementary material II. \cite{sup-mat-II} Moreover, in Sup. Mat. II \cite{sup-mat-II}, we have computed the parameters for the P\"oschl-Teller potential. However, this case has been done numerically, not analytically, and this is why it is not presented here.
\subsection{Hard-sphere potential}
The hard-sphere potential is one of the simplest potential models.  It only has 
one parameter, which is the size of the repulsive core. We point out that the 
formulas derived in Sec. III do not work for 
this particular potential due to its divergent behavior at the core.
The well-known hard-sphere potential is given by
\begin{equation}
V(r) = \begin{cases}
\infty &\mbox{if}\quad r\le R \\
0      &\mbox{if}\quad r > R             \ .
\end{cases}
\end{equation}
The wave function is zero for $r \le R$ and 
\begin{equation}
    u_l(r>R)=\frac{Nk^{-l}}{A_l}r\bigg[j_l(kr)-n_l(kr)\tan{\delta}\bigg]
\end{equation}
for $r>R$. We recall that $k=\sqrt{2\mu E/\hbar^2}$. The scattering length and effective range for any angular momentum $l$ are:
\begin{eqnarray}
     a_l & = & R   \\
    r_l  & = & -\bigg(\frac{1}{(2l+3)}+\frac{1}{(2l-1)}\bigg)R  \ ,
\end{eqnarray}
with $a_l$ being the scattering 
length and $r_l$ the effective range (hereafter, we have eliminated the 
superscript ``eff" in the effective range to simplify the notation).
As we can see, for hard spheres the scattering length is $R$ at any $l$, whereas the effective range is also proportional to $R$ with a value that is positive only for $l=0$, $r_0= 2/3 R$.

\subsection{Soft-sphere potential}
\label{seccioSSP}
In contrast to the hard-sphere interaction, the soft-sphere potential takes into account that the core, although being repulsive, may be penetrated. This correction results in the model potential having two parameters: the strength of the potential and the core size. The soft-sphere potential is given by
\begin{equation}
V(r) = \begin{cases}
V_0 &\mbox{if}\quad r\le R \\
0      &\mbox{if}\quad r > R             \ .
\end{cases}
\end{equation}
For $r \le  R$ , the Schr\"odinger equation is:
\begin{equation}
     u_l''(r)+\bigg(\overline{k}^2-\frac{l(l+1)}{r^2}\bigg)u_l(r)=0\quad \text{where}\quad \overline{k}^2=\frac{2\mu}{\hbar^2}(E-V_0) \ ,
\end{equation}
with $\overline{k}=\sqrt{k^2-k_0^2}$, a complex number for energies $E<V_0$. We have defined $k_0$ as $\sqrt{2\mu V_0/\hbar^2}$. 
In order to have a real function, we need to multiply the Bessel function with an imaginary argument by $i^{-l}$:
\begin{equation}
     u_l(r)=Nrj_l(\overline{k}r)i^{-l} \ .
\end{equation}
For $r > R$ and $k \to 0$, we know the asymptotic solution (see Sec II) is:
\begin{equation}
    u_l(r,k=0)=r^{l+1}-a_l^{2l+1}r^{-l}  \ .
\end{equation}
By imposing that the wave function is continuous at $r=R$, we can find the normalization constant  $N$:
\begin{equation}
     N=\frac{R^{l+1}-a_l^{2l+1}R^{-l}}{Ri^{-l}j_l(ik_0R)} \ .
\label{eqN1}
     \end{equation}
After applying the formulas we have derived in Sec. III, we can obtain the scattering length $a_l$:
\begin{equation}     
a_l^{2l+1}=R^{2l+1}\frac{k_0Ri^{-l-1}j_{l+1}(ik_0R)}{(2l+1)i^{-l}
j_l(ik_0R)+k_0Ri^{-l-1}j_{l+1}(ik_0R)} \ .
\label{asoft}
\end{equation}
The effective range for any $l$ is:
\begin{equation}   
r_l=\frac{2}{(2l+1)a_l^{2l+2}}\bigg\{\frac{R^{2l+3}}{2l+3}-a_l^{2l+1}R^2+\frac{ 
a_l^{4l+2}R^{1-2l}}{1-2l}-N^2(i^{-l})^2\frac{1}{2}R^3\Big(j_l^2(ik_0R)-j_{l-1} 
(ik_0R)j_{l+1}(ik_0R))\Big)\bigg\} \ .
\label{resoft}
\end{equation}
From the general results (\ref{asoft}) and (\ref{resoft}), we can find the 
expressions for $l=0$ and $l=1$ by introducing the Bessel functions 
explicitly. The scattering lengths are:
\begin{eqnarray}
    a_0 & = & R\bigg[1-\frac{\tanh{(k_0R)}}{k_0R}\bigg] 
    \label{a0} \ , \\
    a_1^3 & =& R^3\bigg[\frac{3+(k_0R)^2-3k_0R\coth{(k_0R)}}{(k_0R)^2}\bigg] \ ,
\label{a1}
    \end{eqnarray}    
and the effective ranges are:
\begin{eqnarray}
    r_0 
& = & R-R\frac{(k_0R)^2}{3(k_0R)^2\bigg(1-\cfrac{\tanh{(k_0R)}}{k_0R}\bigg)^2} 
+R\frac{1}{(k_0R)^2\bigg(1-\cfrac{\tanh{(k_0R)}}{k_0R}\bigg)} \ ,
\label{r0}\\
    r_1 
& = & -3k_0a_1^2\frac{-2k_0R(-5+(k_0R)^2)+2k_0R(10+(k_0R)^2)\cosh{(2k_0R)}
-5(3+2(k_0R)^2)\sinh{(2k_0R)}}{10\bigg(-3k_0R\cosh{(k_0R)+(3+(k_0R)^2)\sinh{
(k_0R)}}\bigg)^2} \ .
\label{r1}
\end{eqnarray}

\begin{figure}[H]
    \centering
    \begin{subfigure}[b]{0.49\textwidth}
        \centering
        \includegraphics[width=\textwidth]{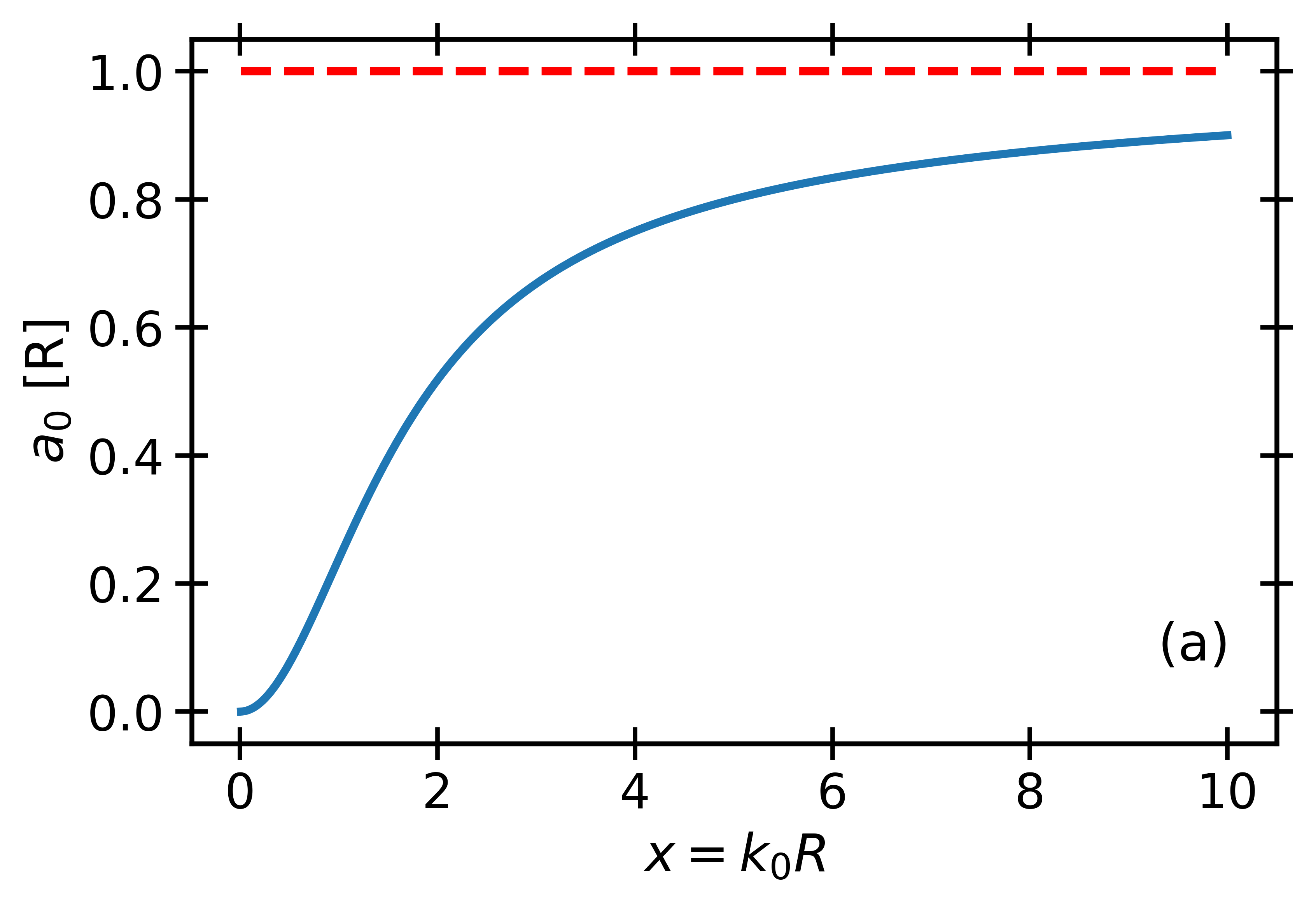}
    \end{subfigure}
    \begin{subfigure}[b]{0.49\textwidth}
        \centering
        \includegraphics[width=\textwidth]{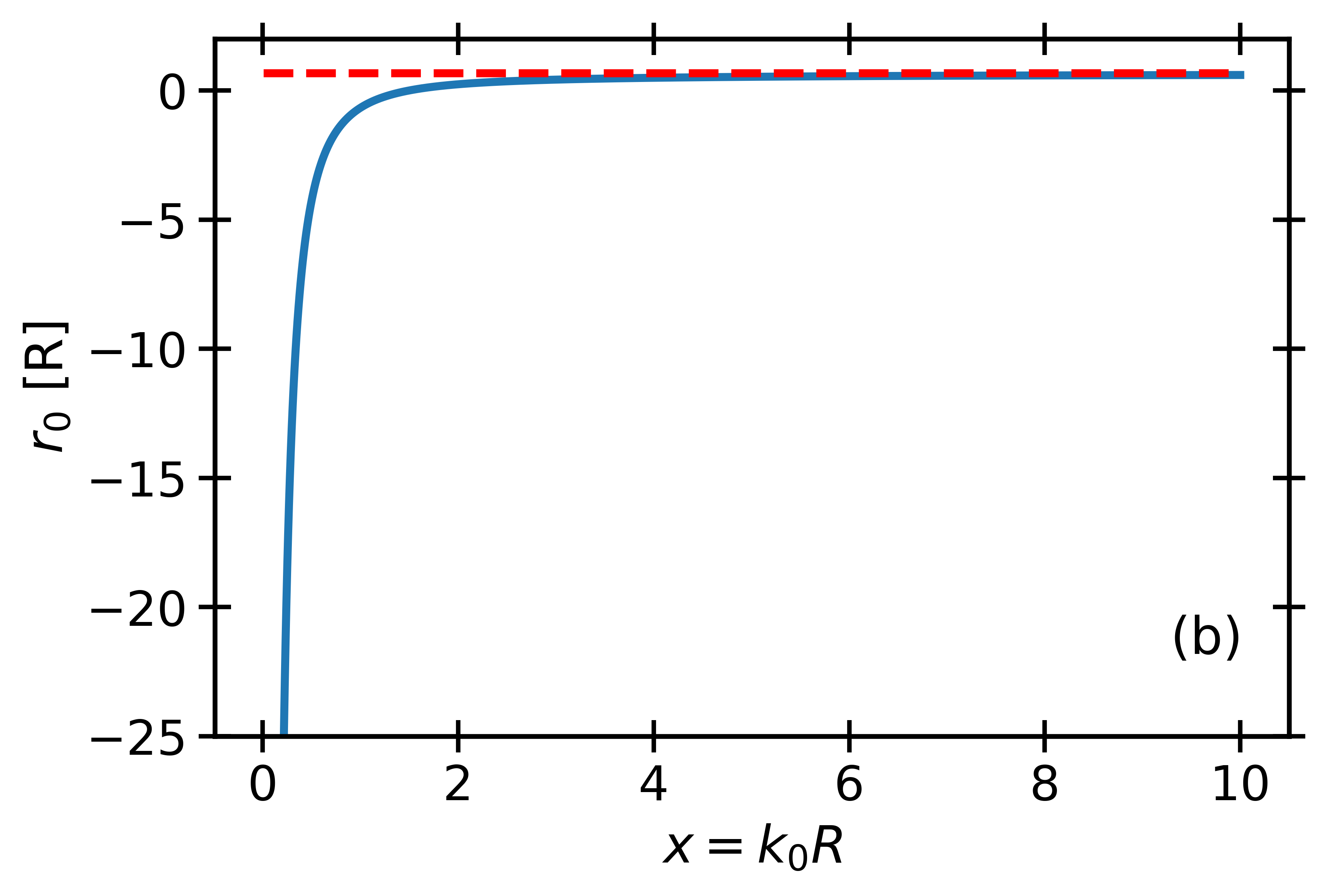}
    \end{subfigure}
    \\
    \begin{subfigure}[b]{0.49\textwidth}
        \centering
        \includegraphics[width=\textwidth]{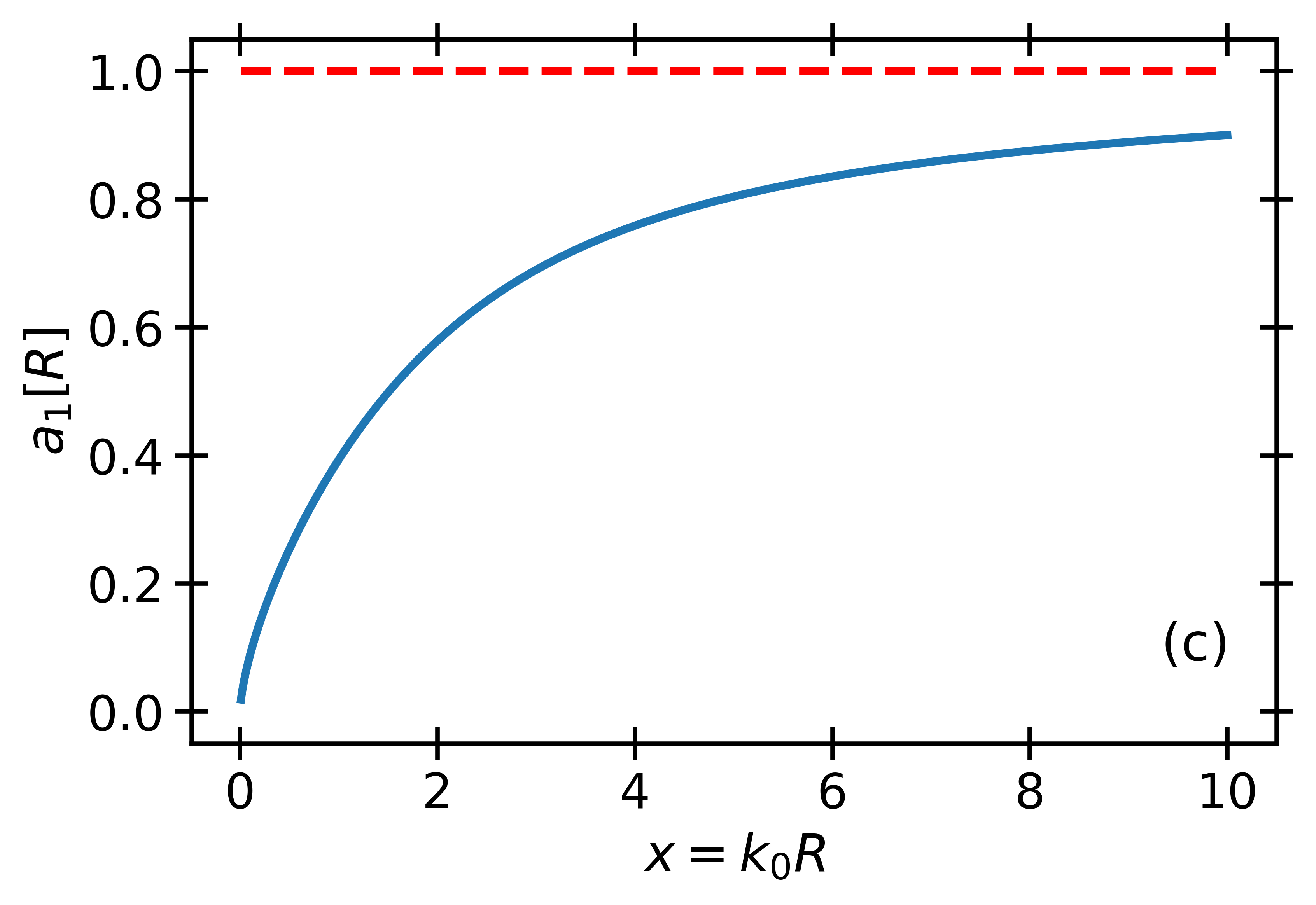}
    \end{subfigure}
    \begin{subfigure}[b]{0.49\textwidth}
        \centering
        \includegraphics[width=\textwidth]{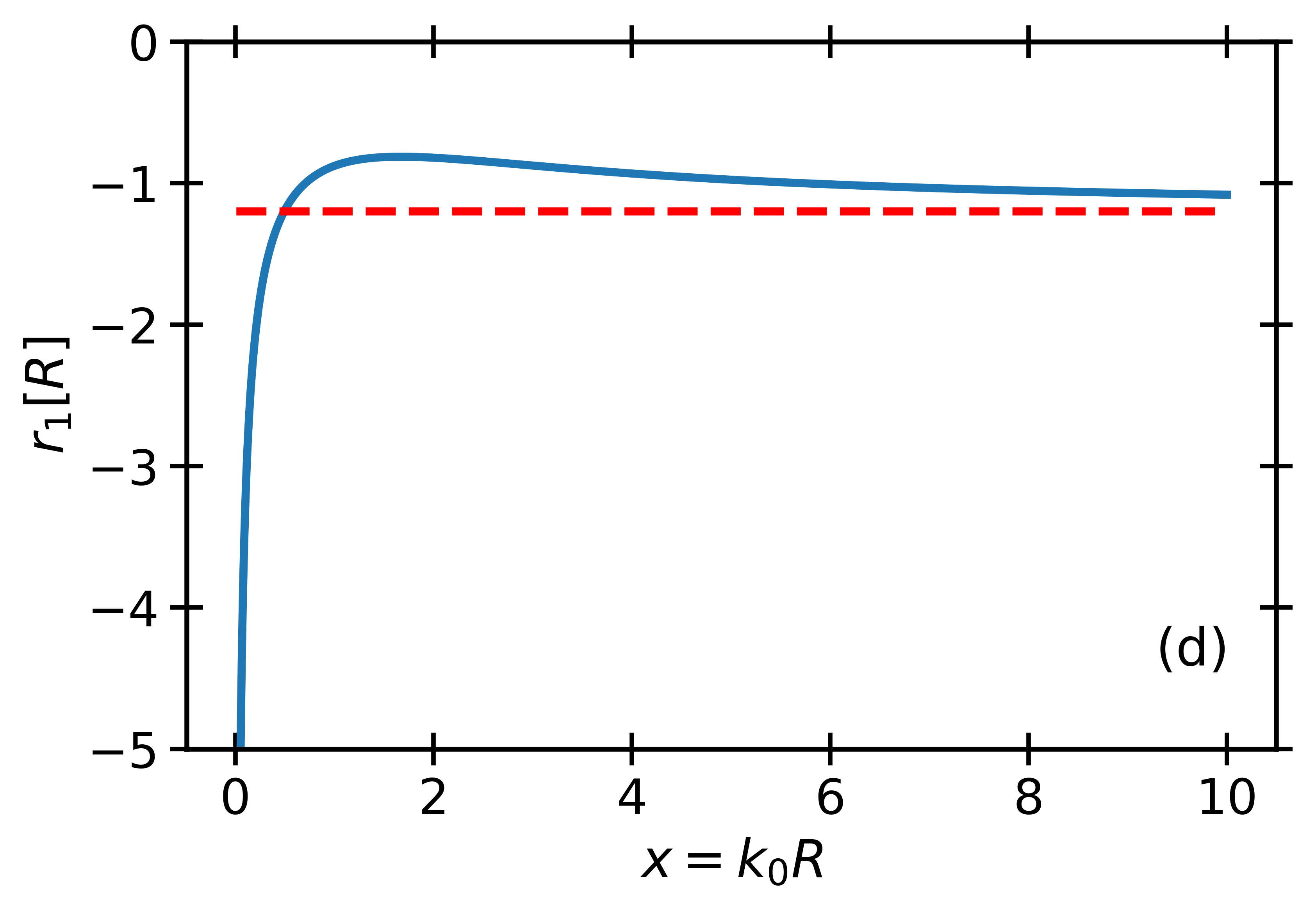}
    \end{subfigure}
    \caption{\footnotesize{ $a_0$, $r_0$, $a_1$ and $r_1$ for a 
soft-sphere potential as a function of $k_0 R$. The red dashed lines correspond to the hard-sphere limit.}}
\label{fig.soft}
\end{figure}

In Fig. \ref{fig.soft}, we plot the functions (\ref{a0},\ref{a1},\ref{r0},\ref{r1}).  
The scattering lengths $a_0$ and $a_1$ are zero when $k_0R$ is zero, and are 
equal to $R$ (hard-spheres solution) when $k_0R$ becomes large. In contrast, $r_0$ 
and $r_1$ diverge for $k_0R$ equal to zero, and when $k_0R$ increases we 
recover the value of hard-spheres, which is $2R/3$ and $-6R/5$, respectively.

An alternate way of finding the scattering parameters of the soft-sphere potential is shown in the Appendix of the Supplemental Material II \cite{sup-mat-II}.

\subsection{Spherical well potential}
\label{seccioSSA}
\begin{figure}[H]
    \centering
    \begin{subfigure}[b]{0.49\textwidth}
        \centering
        \includegraphics[width=\textwidth]{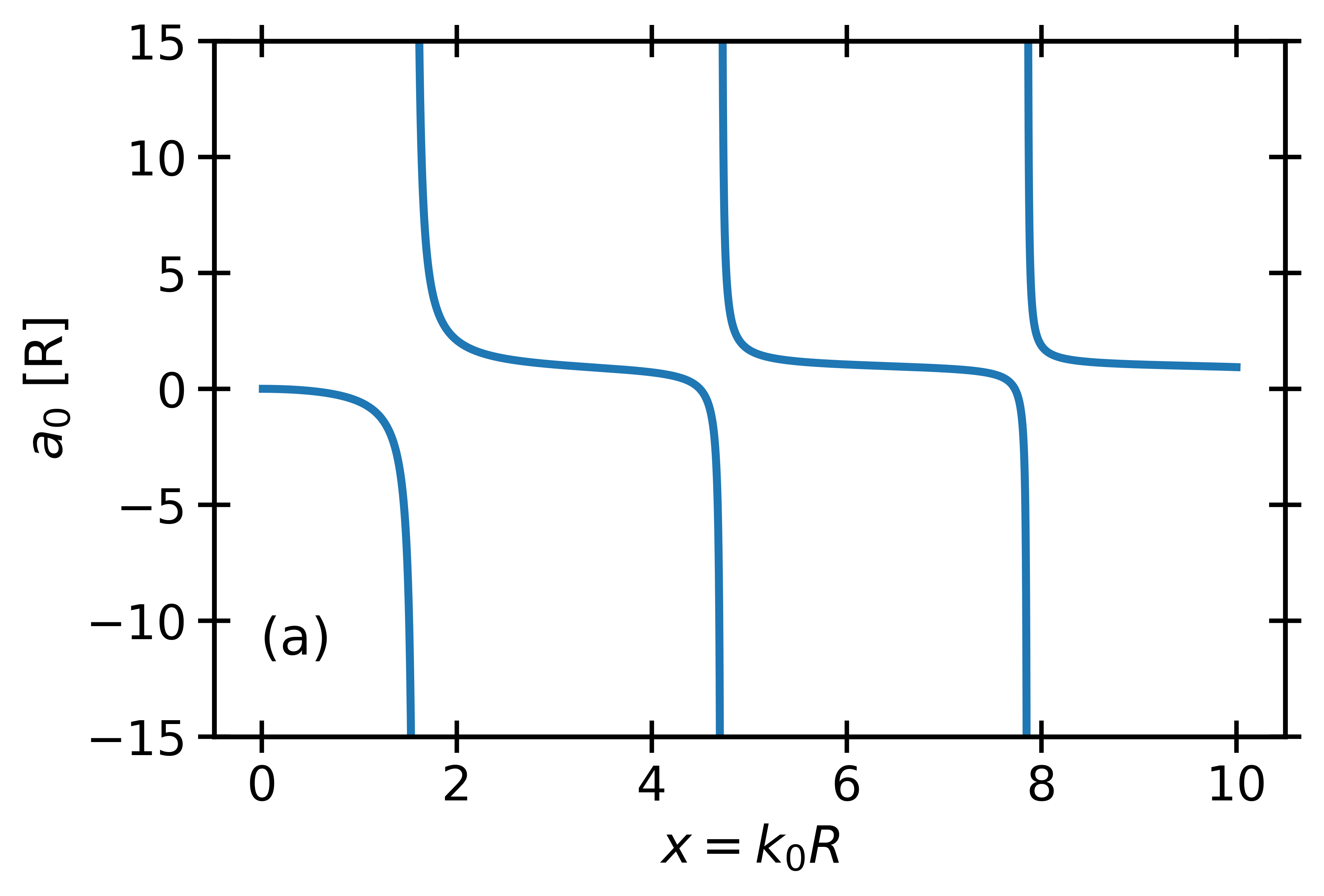}
    \end{subfigure}
    \begin{subfigure}[b]{0.49\textwidth}
        \centering
        \includegraphics[width=\textwidth]{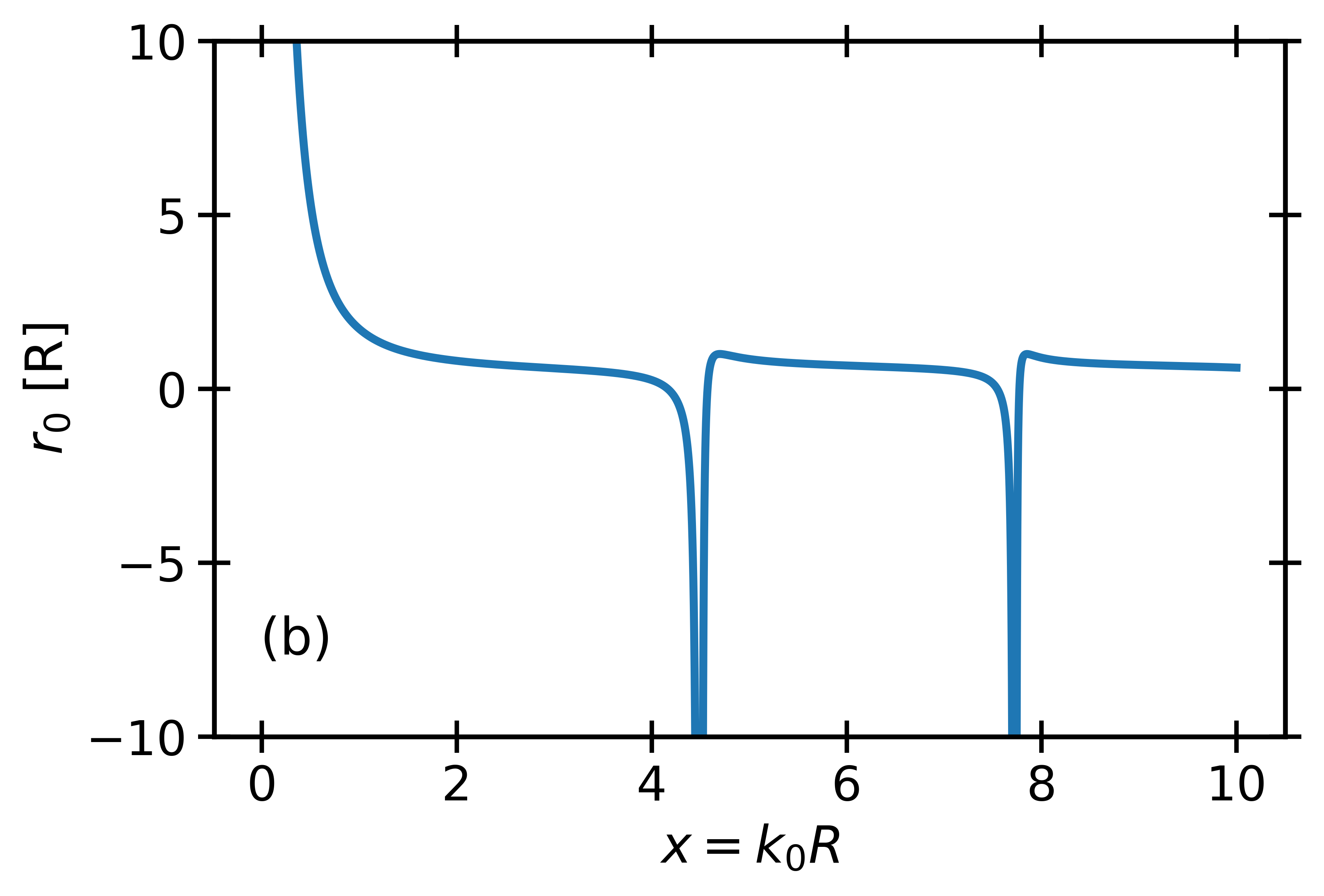}
    \end{subfigure}
    \\
    \begin{subfigure}[b]{0.49\textwidth}
        \centering
        \includegraphics[width=\textwidth]{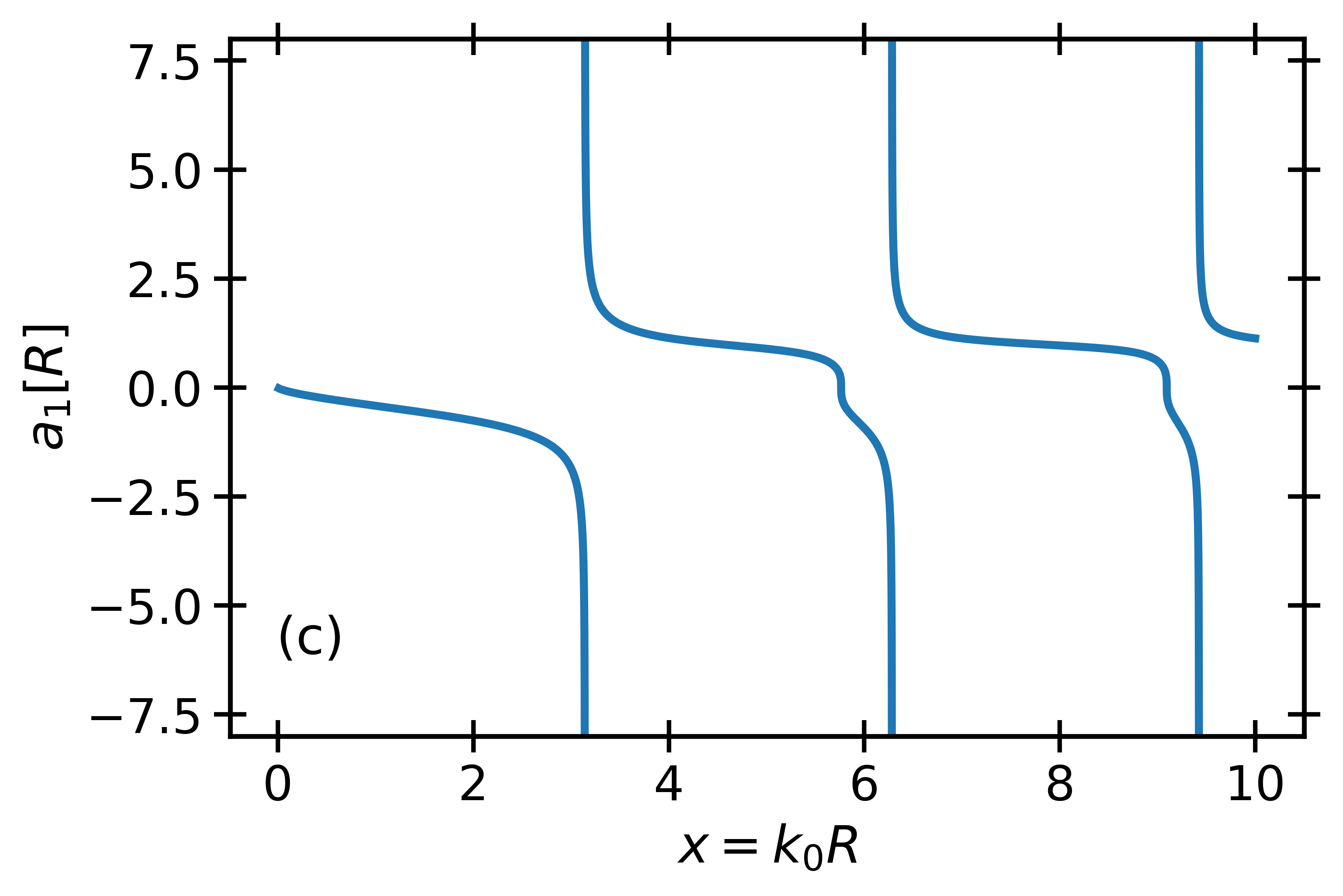}
    \end{subfigure}
    \begin{subfigure}[b]{0.49\textwidth}
        \centering
        \includegraphics[width=\textwidth]{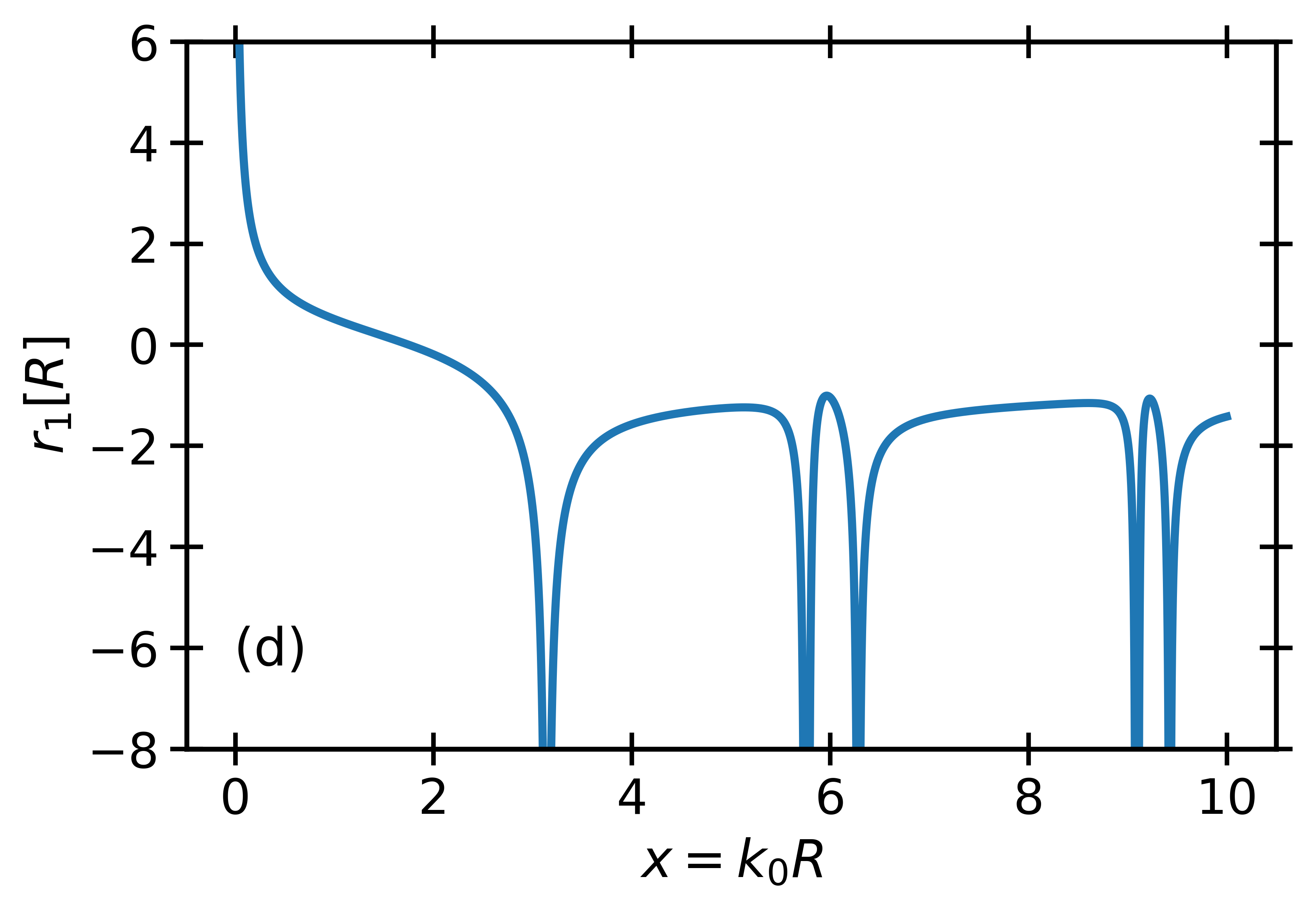}
    \end{subfigure}
    \caption{\footnotesize{Representation of $a_0$, $r_0$, $a_1$ and $r_1$ for a 
spherical well.}}
\label{fig.squarewell}
\end{figure}

In the previous Section, we solved the soft-sphere 
model potential, where we have a potential barrier. The spherical well potential is the same idea of the soft-sphere potential but taking it as an attractive potential,
\begin{equation}
V(r) = \begin{cases}
-V_0 &\mbox{if}\quad r\le R \\
0      &\mbox{if}\quad r > R             \ .
\end{cases}
\end{equation}
For the spherical well, where the potential is negative, we can use the expressions that 
we have obtained for soft-spheres by simply changing $k_0$ to $-ik_0$. Note that since $k_0$ depends on the square root of $V_0$, a sign change in $V_0$ results in $k_0$ becoming imaginary. If we apply this change in sign to $V_0$, then 
we arrive easily to the general formulas for any $l$. For the lowest partial 
waves $l=0$ and $l=1$, the scattering lengths are:
\begin{eqnarray}
    a_0 & =& R\bigg[1-\frac{\tan{(k_0R)}}{k_0R}\bigg] \ ,  \\
    a_1^3 & =& R^3\bigg[\frac{(k_0R)^2-3+3k_0R\cot{(k_0R)}}{(k_0R)^2}\bigg] \ ,
\end{eqnarray}
and the effective ranges are:
\begin{eqnarray}
r_0 & = & 
R-R\frac{(k_0R)^2}{3(k_0R)^2\bigg(1-\cfrac{\tan{(k_0R)}}{k_0R}\bigg)^2} 
-R\frac{1}{(k_0R)^2\bigg(1-\cfrac{\tan{(k_0R)}}{k_0R}\bigg)} \ , \\
    r_1 
& = & -3k_0a_1^2\frac{2k_0R(5+(k_0R)^2)+2k_0R(10-(k_0R)^2)\cos{(2k_0R)}
-5(3-2(k_0R)^2)\sin{(2k_0R)}}{10\bigg(3k_0R\cos{(k_0R)-(3-(k_0R)^2)\sin{(k_0R)}}
\bigg)^2} \ .
\end{eqnarray}
As we discussed in Sec. IV, in an attractive potential and at  certain values of the potential strength, the scattering length diverges (Feshbach resonance). That means that a bound state appears. For the spherical well potential, the resonance for $l=0$ occurs at $k_0R=(2n-1)\pi/2$ and for l=1 at $k_0R=n\pi$, with $n\geq1$. In Fig. \ref{fig.squarewell}, we plot the scattering lengths and effective 
ranges for $l=0$ and $l=1$.  We observe the predicted divergences: $a_0$ diverges at 
$k_0R=$ $\pi$/2, 3$\pi$/2 and 5$\pi$/2. $a_1$ diverges at $k_0R=$ $\pi$, 2$\pi$ and 3$\pi$.
If one looks to the shape of the effective ranges, one can see some divergences 
appearing. As we proved in the end of Sec. IV, the divergences of the effective ranges are related to the 
divergences of the scattering length and to its zero value. 
In particular, for $r_0$ the 
divergences occur  when $a_0=0$, while for $r_1$, we see divergences 
coming from both limits, that is when  $a_1=0$ and  
$a_1=\pm\infty$. From that, we can also understand why we were not seeing a 
divergence in the effective range, $r_0$, near the region of the first divergence of the 
scattering length. This is because, before the first divergence of $a_0$, $a_0$ 
is already negative and we do not see a change in the sign (hence, passing 
through zero) until we go to the second divergence of $a_0$.

\subsection{Well-barrier potential}
This model potential is more involved than the previous ones. It is a combination of a spherical-well potential and a soft-sphere (barrier) potential. Due to this combination, with this model potential we have four parameters at our disposal: the strengths and the sizes of the spherical-well and the barrier. This model potential is very well suited when working with perturbative expansions of many-body systems that contain the three most important scattering parameters ($a_0$, $a_1$ and $r_0$) as it more easily fits them than in the previous model potentials. It is given by
\begin{equation}
V(r) = \begin{cases}
-V_1 &\mbox{if}\quad r\le R_1 \\
V_2  &\mbox{if}\quad R_1 < r \le R_2 \\
0      &\mbox{if}\quad r > R_2             \ .
\end{cases}
\end{equation}
In the internal region we have a 
well potential, in which the wave function will be the regular Bessel function 
with a real argument:
\begin{equation}
    u_l^{(0)}(r)=N_1rj_l(k_1r) \quad \text{if}\quad r \le R_1,\quad \text{where}\quad k_1=\sqrt{2\mu V_1/\hbar^2}   \ .
\end{equation}
In  the second region, the solution is a combination of Bessel 
functions with imaginary arguments, just like in the barrier potential:
\begin{equation}
    u_l^{(0)}(r)=N_2r\Big[j_l(ik_2r)+M_2n_l(ik_2r)\Big]\quad \text{if}\quad R_1<r \le R_2,\quad \text{where}\quad k_2=\sqrt{2\mu V_2/\hbar^2} \ 
.
\end{equation}
And finally, beyond $R_2$, we have the 
long-distance solution:
\begin{equation}
    u_l^{(0)}(r)=r^{l+1}-a_l^{2l+1}r^{-l} \quad \text{if}\quad r>R_2 \ .
\end{equation}
We need the constants of integration:
\begin{equation}
N_1=N_2\frac{j_l(ik_2R_1)+M_2n_l(ik_2R_1)}{j_l(k_1R_1)}\quad;\quad N_2=\frac{R_2^{l+1}-a_l^{2l+1}{R_2}^{-l}}{R_2\big[j_l(ik_2R_2)+M_2n_l(ik_2R_2)\big]}
\end{equation}
\begin{equation}
    \text{and}\quad M_2=-\frac{ik_2R_1j_l(k_1R_1)j_{l+1}(ik_2R_1)-k_1R_1j_l(ik_2R_1)j_{l+1}(k_1R_1)}{ik_2R_1j_l(k_1R_1)n_{l+1}(ik_2R_1)-k_1R_1n_l(ik_2R_1)j_{l+1}(k_1R_1)} \ .
\end{equation}
The $l$-wave scattering length is:
\begin{equation}
   a_l^{2l+1}=\frac{R_2^{2l+1}
\displaystyle\int_0^{R_2}U(r)r^{l+1}u_l^{(0)}(r)dr}{(2l+1)R_2^{l}u_l^{(0)}(R_2)+\displaystyle\int_0^{R_2}U(r)r^{l+1}u_l^{(0)}(r)dr} \ ,
\label{eqal}
\end{equation}
where $u_l^{(0)}(R_2)$ is $N_2R_2\Big[j_l(ik_2R_2)+M_2n_l(ik_2R_2)\Big]$ and the integrals are:
\begin{equation}
    \int_0^{R_2}U(r)r^{l+1}u_l^{(0)}(r)dr=-N_1k_1R_1^{l+2}j_{l+1}(k_1r)-iN_2k_2\bigg[r^{l+2}j_{l+1}(ik_2r)+M_2(-1)^lr^{l+2}j_{-l-2}(ik_2r)\bigg]_{R_1}^{R_2} \ .
\end{equation}
Finally, the effective range can be written as
\begin{equation}
    r_l=\frac{2}{(2l+1)a_l^{2l+2}}\bigg\{\frac{R_2^{2l+3}}{2l+3}-a_l^{2l+1}R_2^2+\frac{a_l^{4l+2}R_2^{1-2l}}{1-2l}-N_1^2I_{jj}(k_1R_1)-N_2^2\bigg[I_{jj}(ik_2r)+M_2^2I_{nn}(ik_2r)+2M_2I_{jn}(ik_2r)\bigg]_{R_1}^{R_2}\bigg\} \ .
    \label{rewbp}
\end{equation}
The functions $I_{jj}(kr)$, $I_{nn}(kr)$ and $I_{jn}(kr)$ in Eq. (\ref{rewbp}) are:
\begin{equation}
    I_{jj}(kr)=\int r^2j_l^2(kr)dr=\frac{r^3}{2}\big(j_l^2(kr)-j_{l-1}(kr)j_{l+1}(kr)\big) \ ,
\end{equation}
\begin{equation}
    I_{nn}(kr)=\int r^2n_l^2(kr)dr=\frac{r^3}{2}\big(n_l^2(kr)-n_{l-1}(kr)n_{l+1}(kr)\big) \ ,
\end{equation}
\begin{equation}
\begin{aligned}
    I_{jn}(kr)=\int r^2j_l(kr)n_l(kr)dr=\frac{(-1)^lr^3}{2}j_{-l}(kr)j_{l-1}(kr)-\frac{(-1)^lr^3}{2}j_{-l-1}(kr)j_{l}(kr)\\
    -\frac{(1+2l)(-1)^lr^2}{4k}j_{-l}(kr)j_{l}(kr)+\frac{(1+2l)(-1)^lr^2}{4k}j_{-l-1}(kr)j_{l-1}(kr) \ .
\end{aligned}
\end{equation}

The scattering length for $l=0$ is:
\begin{equation}
    a_0=\frac{\cosh{(k_2R_{12})}[-k_1k_2R_2\cos{(k_1R_1)+k_2\sin{(k_1R_1)}}]+\sinh{(k_2R_{12})}[k_2^2R_2\sin{(k_1R_1)}-k_1\cos{(k_1R_1)}]}{-k_1k_2\cos{(k_1R_1)}\cosh{(k_2R_{12})}+k_2^2\sin{(k_1R_1)}\sinh{(k_2R_{12})}} \ ,
\end{equation}
with $R_{12}=R_1-R_2$.
For $a_1$ and $r_0$, the expressions are complicated but still manageable, and 
hence, we can split them into parts. For  $a_1$  we get:
\begin{equation}
a_1^3=\frac{p_1+p_2}{d_1} \ ,
\end{equation}
with
\begin{eqnarray}
    p_1 
& = & \cosh{(k_2R_{12})}\{3k_1k_2^3R_1R_2^2\cos{
(k_1R_1)+k_2R_2[-3k_2^2R_2+k_1^2(3R_1-3R_2+k_2^2R_1R_2^2)]\sin{(k_1R_1)}}\} \ ,\hspace{
1.85cm } \\
    p_2 
& = & \sinh{(
k_2R_{12})}\{k_1k_2^2R_1R_2[3+k_2^2R_2^2]\cos{(k_1R_1)} \nonumber \\
&&-R_2[3k_2^2+3k_1^2+k_2^4R_2^2+k_1^2k_2^2R_2(-3R_1+R_2)]\sin{(k_1R_1)}\} \ , \\
    d_1 
& = & k_1^2k_2^3R_1\sin{(k_1R_1)}\cosh{(k_2R_{12})}+[k_1k_2^4R_1\cos{
(k_1R_1) }-k_2^2(k_1^2+k_2^2)\sin{(k_1R_1)}]\sinh{(k_2R_{12})} \ .
\end{eqnarray}
Similarly, for  $r_0$:
\begin{equation}
 r_0=\frac{p_3+p_4+p_5}{d_2} \ ,
\end{equation}
with
\begin{eqnarray}
    p_3 
& = & 
2k_1k_2\cosh{(2k_2R_{12})}\{R_2(3+k_2^2R_2^2)[k_1^2+k_2^2+(k_1^2-k_2^2)\cos{ 
(2k_1R_1)}]-3k_1(1+2k_2^2R_2^2)\sin{(2k_1R_1)}\} \ , \hspace{1cm}\\
    p_4 
& =  & 
2k_2\{-3k_1(k_1^2+k_2^2)R_1+k_1(k_1^2-k_2^2)k_2^2R_2^3+(k_1^2+k_2^2)[k_1(-3R_1+ k 
_2^2R_2^3)\cos{(2k_1R_1)}+3\sin{(2k_1R_1)}]\} \ ,\\
    p_5 
& = & k_1\{3(1+2k_2^2R_2^2)[k_1^2+k_2^2+(k_1^2-k_2^2)\cos{(2k_1R_1)}
]-4k_1k_2^2R_2(3+K_2^2R_2^2)\sin{(2k_1R_1)}\}\sinh{(2k_2R_{12})} \ ,\\          
d_2 & = 
& 12k_1k_2\{k_2\cosh{(k_2R_{12})}[k_1R_2\cos{(k_1R_1)}-\sin{(k_1R_1)}] 
\nonumber \\
       && +[k_1\cos{(k_1R_1)}-k_2^2R_2\sin{(k_1R_1)}]\sinh{(k_2R_{12})}\}^2 \ .
\end{eqnarray}
The formulas found for $a_0$ and $r_0$ coincide with the ones in Ref. \cite{scawellbarrier}.
The analytic expressions for the rest of parameters are too large and we do not 
show them in this work.


\section{Conclusions}
In this work, we have re-analyzed the scattering problem to find expressions for obtaining the scattering length and effective range for any angular momentum value $l$ as long as the potential decays faster than $1/r^n$ with $n>2l+3$ for each partial wave at infinity. We have first worked out the procedure that leads to integral formulas allowing the calculation of the scattering parameters for any angular momentum. The procedure consists of relating two apparently different formulas that calculate the phase shift. After lengthy manipulations, one obtains the expressions for the scattering parameters. Also, we discuss the Feshbach resonance and how the scattering parameters behave around it, with special emphasis on the importance of the resonance in the study of ultracold quantum gases.

As model examples, we have obtained the exact expressions of the scattering length and the effective range for any angular momentum $l$ for the cases of hard-sphere, soft-sphere, spherical well and well-barrier potentials. As the more useful parameters are $a_0$, $a_1$ and $r_0$, we also particularize the general formulas for these cases.

The results obtained for the analyzed model potentials can be useful for researchers working on perturbative expansions used in many-body theory when the interaction strength is small \cite{scaguardiola,genecoef,Bishop}. They can also be useful when working the inverse problem, where one tries to build potentials which fulfill 
some a-priori known low-energy scattering parameters. Finally, Sections II, III and IV can be used as a a guide for students studying scattering theory, and the model problems in Sec. V can be used as hands-on applied exercises.

\begin{acknowledgments}
This work has been supported by AEI (Spain) under grant No. 
PID2020-113565GB-C21.   We also acknowledge financial support from Secretaria d'Universitats i Recerca del Departament d'Empresa i Coneixement de la Generalitat de Catalunya, co-funded by the European Union Regional Development Fund within the ERDF Operational Program of Catalunya (project QuantumCat, ref. 001-P-001644).
\end{acknowledgments}


\bibliography{apssamp}

\appendix

\end{document}


\preprint{}

\title{Low-energy scattering parameters:  A theoretical derivation of the 
effective range and scattering length for arbitrary angular momentum\\ Supplementary material I}

\author{Jordi Pera}
\author{Jordi Boronat}%
\affiliation{%
Departament de Física, Campus Nord B4-B5, Universitat Politècnica de Catalunya, 
E-08034 Barcelona, Spain
}%


\maketitle

\section{Potential substitution and integration by parts}
\label{appendix:a}
We want to manipulate the coefficient $c_2$,
\begin{equation}    
c_2=\frac{2}{2l+1}\int_0^{\infty}U(r)\bigg(r^{l+1}u_l^{(2)}-\frac{r^{l+3}u_l^{(0)}}{
2(2l+3)}\bigg)dr \ .
\end{equation}
We will use the differential equations for $u_0$ and $u_2$ to cancel  
the potential, and afterwards we will apply integration by parts to 
get rid of the second derivatives.

For $u_l^{(0)}$,
\begin{equation}
    U(r)u_l^{(0)}=u_l^{(0)''}-\frac{l(l+1)}{r^2}u_l^{(0)}
\end{equation}
\begin{equation}
    \int_0^{\infty}U(r)u_l^{(0)}r^{l+3}dr=r^{l+3}u_l^{(0)'}\Big|_0^{\infty}-(l+3)r^{l+2}u_l^{(0)}\Big|_0^{\infty}+\int_0^{\infty}(l+3)(l+2)r^{l+1}u_l^{(0)}dr-\int_0^{\infty}l(l+1)r^{l+1}u_l^{(0)}dr \ .
\end{equation}
For $u_l^{(2)}$,
\begin{equation}
    U(r)u_l^{(2)}=u_l^{(2)''}-\frac{l(l+1)}{r^2}u_l^{(2)}+u_l^{(0)}
\end{equation}
\begin{equation}
    \int_0^{\infty}U(r)u_l^{(2)}r^{l+1}dr=r^{l+1}u_l^{(2)'}\Big|_0^{\infty}-(l+1)r^lu_l^{(2)}\Big|_0^{\infty}+\int_0^{\infty}r^{l+1}u_l^{(0)}dr \ .
\end{equation}
Substituting everything into $c_2$,
\begin{equation*}
    c_2=\frac{2}{2l+1}\bigg\{r^{l+1}u_l^{(2)'}\Big|_0^{\infty}-(l+1)r^lu_l^{(2)}\Big|_0^{\infty}+\int_0^{\infty}r^{l+1}u_l^{(0)}dr
\end{equation*}
\begin{equation*}
    +\frac{1}{2(2l+3)}\bigg((l+3)r^{l+2}u_l^{(0)}\Big|_0^{\infty}-r^{l+3}u_l^{(0)'}\Big|_0^{\infty}-2(2l+3)\int_0^{\infty}r^{l+1}u_l^{(0)}dr\bigg)\bigg\} \ .
\end{equation*}

Further rearranging leads to the expression (Eq. (43)) that we 
have in Sec. III of Ref. \cite{Part-I}, that is,
\begin{equation}
    \frac{(2l+1)c_2}{2}=r^{l+1}u_l^{(2)'}\Big|_0^{\infty}-(l+1)r^lu_l^{(2)}\Big|_0^{\infty}+\frac{1}{2(2l+3)}\bigg\{(l+3)r^{l+2}u_l^{(0)}\Big|_0^{\infty}-r^{l+3}u_l^{(0)'}\Big|_0^{\infty}\bigg\} \ .
\end{equation}

\section{Substitution of $u_l^{(0)}$, $u_l^{(0)'}$ and $u_l^{(2)}$}
\label{appendix:b}
The expression for $c_2$ is the following,
\begin{equation}
    \frac{(2l+1)c_2}{2}=\lim_{r\rightarrow\infty}\bigg\{\frac{(l+3)r^{l+2}u_l^{(0)}}{2(2l+3)}-\frac{r^{l+3}u_l^{(0)'}}{2(2l+3)}-\frac{u_l^{(2)}}{u_l^{(0)}}[(l+1)r^lu_l^{(0)}-r^{l+1}u_l^{(0)'}]-\frac{r^{l+1}}{u_l^{(0)}}\int_0^{r}{u_l^{(0)}}^2(r)dr\bigg\} \ .
\end{equation}
We recall the expressions for $u_l^{(0)}$, $u_l^{(0)'}$ and $u_l^{(2)}$:
\begin{gather}
    u_l^{(0)}=r^{l+1}-a_l^{2l+1}r^{-l}\quad,\quad u_l^{(0)'}=(l+1)r^l+la_l^{2l+1}r^{-l-1}\\
    u_l^{(2)}=-\frac{r^{l+3}}{2(2l+3)}-a_l^{2l+1}\frac{r^{2-l}}{2(2l-1)}-\frac{1}{2}a_l^{2l+2}r_l^{\text{eff}}r^{-l}
\end{gather}
The next step is the substitution of $u_l^{(0)}$, $u_l^{(0)'}$, and $u_l^{(2)}$ 
into the expression of $c_2$. We will do this by parts.

The two first terms are
\begin{equation*}
    \frac{(l+3)r^{l+2}u_l^{(0)}}{2(2l+3)}-\frac{r^{l+3}u_l^{(0)'}}{2(2l+3)}=\frac{(l+3)r^{l+2}}{2(2l+3)}\Big(r^{l+1}-a_l^{2l+1}r^{-l}\Big)-\frac{r^{l+3}}{2(2l+3)}\Big((l+1)r^l+la_l^{2l+1}r^{-l-1}\Big)
\end{equation*}
\begin{equation}
    =\frac{r^{2l+3}}{2l+3}-\frac{a_l^{2l+1}r^2}{2}   \ .
\end{equation}

The next term is:
\begin{equation*}
    \frac{u_l^{(2)}}{u_l^{(0)}}[(l+1)r^lu_l^{(0)}-r^{l+1}u_l^{(0)'}]=\frac{u_l^{(2)}}{u_l^{(0)}}\Big[(l+1)r^l\Big(r^{l+1}-a_l^{2l+1}r^{-l}\Big)-r^{l+1}\Big((l+1)r^l+la_l^{2l+1}r^{-l-1}\Big)\Big]
\end{equation*}
\begin{equation*}
    =-\frac{u_l^{(2)}}{u_l^{(0)}}(2l+1)a_l^{2l+1}
\end{equation*}
As we are in the limit of long distances, we can use the following approximation,
\begin{equation}    
\frac{1}{u_l^{(0)}}=\frac{1}{r^{l+1}-a_l^{2l+1}r^{-l}}=\frac{1}{r^{l+1}}\frac{1}
{1-a_l^{2l+1}r^{-2l-1}}\approx\frac{1}{r^{l+1}}\bigg(1+a_l^{2l+1}r^{-2l-1}+a_l^{
4l+2}r^{-4l-2}\bigg) \ .
\end{equation}
Therefore, we have
\begin{equation*}
-\frac{u_l^{(2)}}{u_l^{(0)}}(2l+1)a_l^{2l+1}\approx\bigg(\frac{r^{l+3}}{2(2l+3)}
+a_l^{2l+1}\frac{r^{2-l}}{2(2l-1)}+\frac{1}{2}a_l^{2l+2}r_lr^{-l}\bigg)\frac{1}{
r^{l+1}}\bigg(1+a_l^{2l+1}r^{-2l-1}+a_l^{4l+2}r^{-4l-2}\bigg)(2l+1)a_l^{2l+1}  
\ .
\end{equation*}
We remember here  that we are working in the limit $r \to \infty$, meaning that  
any term that at infinity is zero is not considered, resulting in
\begin{equation*}
    -\frac{u_l^{(2)}}{u_l^{(0)}}(2l+1)a_l^{2l+1}=(2l+1)a_l^{2l+1}\bigg(\frac{r^2}{2(2l+3)}+\frac{a_l^{2l+1}r^{1-2l}}{2(2l+3)}+\frac{a_l^{4l+2}r^{-4l}}{2(2l+3)}+\frac{a_l^{2l+1}r^{1-2l}}{2(2l-1)}+\frac{a_l^{4l+2}r^{-4l}}{2(2l-1)}\bigg)
\end{equation*}
\begin{equation}
=\frac{2l+1}{2l+3}a_l^{2l+1}\frac{r^2}{2}+\frac{(2l+1)^2}{(2l+3)(2l-1)}a_l^{
4l+2 }\Big(r^{1-2l}+a_l^{2l+1}r^{-4l}\Big)  \ .
\end{equation}

And  the last term is:
\begin{equation*}
\frac{r^{l+1}}{u_l^{(0)}}\int_0^{r}{u_l^{(0)}}^2(r)dr=\frac{r^{l+1}}{r^{l+1}
-a_l^{2l+1}r^{-l}}\int_0^{r}{u_l^{(0)}}^2(r)dr\approx\bigg(1+\frac{a_l^{2l+1}}{
r^{2l+1}}+\frac{a_l^{4l+2}}{r^{4l+2}}+\frac{a_l^{6l+3}}{r^{6l+3}}\bigg)\int_0^{r
}{u_l^{(0)}}^2(r)dr \ . 
\end{equation*}
In order to take the limit, we break the integral in two parts. One goes from 0 to $R$, which is very large but finite. The second term goes from $R$ 
to $r$, with $r$ going to infinity; in this second integral, we can approximate $u_0$ by the limit it takes as $r \to \infty$, 
\begin{equation}    
\frac{r^{l+1}}{u_l^{(0)}}\int_0^{r}{u_l^{(0)}}^2(r)dr=\int_0^{r}{u_l^{(0)}}^2(r)dr+\bigg(\frac{a_l^{2l+1}}{r^{2l+1}}+\frac{a_l^{4l+2}
}{r^{4l+2}}+\frac{a_l^{6l+3}}{r^{6l+3}}\bigg)\bigg(\int_0^{R}{u_l^{(0)}}^2(r)dr+
\int_R^{r}{u_l^{(0)}}^2(r)dr\bigg) \ .
\end{equation}
The first term containing the integral from 0 to $R$ goes to zero because the 
integral has a finite value and the quotients in front of it go to zero at 
infinity:
\begin{equation}
\frac{r^{l+1}}{u_l^{(0)}}\int_0^{r}{u_l^{(0)}}^2(r)dr=\int_0^{r}{u_l^{(0)}}^2(r)dr+\bigg(\frac{a_l^{2l+1}}{r^{2l+1}}+\frac{a_l^{4l+2}
}{r^{4l+2}}+\frac{a_l^{6l+3}}{r^{6l+3}}\bigg)\bigg(\frac{r^{2l+3}}{2l+3}
\Big|_R^r-a_l^{2l+1}r^2\Big|_R^r+\frac{a_l^{4l+2}r^{1-2l}}{1-2l}\Big|_R^r\bigg) 
\ . 
\end{equation}
The terms resulting from the limit $R$ also go to zero because of the same 
reasoning as before. Although $R$ is very large, it is finite. Thus, we only 
care about the upper limit $r$ that is precisely the variable going to infinity:
\begin{equation*}
    \frac{r^{l+1}}{u_l^{(0)}}\int_0^{r}{u_l^{(0)}}^2(r)dr=\int_0^{r}{u_l^{(0)}}^2(r)dr+\bigg(\frac{a_l^{2l+1}}{r^{2l+1}}+\frac{a_l^{4l+2}}{r^{4l+2}}+\frac{a_l^{6l+3}}{r^{6l+3}}\bigg)\bigg(\frac{r^{2l+3}}{2l+3}-a_l^{2l+1}r^2+\frac{a_l^{4l+2}r^{1-2l}}{1-2l}\bigg)
\end{equation*}
\begin{equation}
\approx\int_0^{r}{u_l^{(0)}}^2(r)dr+\frac{a_l^{2l+1}r^2}{2l+3}+a_l^{4l+2}r^{1-2l
}\bigg(\frac{1}{2l+3}-1\bigg)+a_l^{6l+3}r^{-4l}\bigg(\frac{1}{1-2l}+\frac{1}{
2l+3}-1\bigg)  \ .
\end{equation}

Finally, we put everything together:
\begin{equation*}
\frac{(2l+1)c_2}{2}=\frac{(l+3)r^{l+2}u_l^{(0)}}{2(2l+3)}-\frac{r^{l+3}u_l^{(0)'
}}{2(2l+3)}-\frac{u_l^{(2)}}{u_l^{(0)}}[(l+1)r^lu_l^{(0)}-r^{l+1}u_l^{(0)'}]
-\frac{r^{l+1}}{u_l^{(0)}}\int_0^{r}{u_l^{(0)}}^2(r)dr \ . 
\end{equation*}
The first term is:
\begin{equation*}
\frac{(l+3)r^{l+2}u_l^{(0)}}{2(2l+3)}-\frac{r^{l+3}u_l^{(0)'}}{2(2l+3)}=\frac{r^
{2l+3}}{2l+3}-\frac{a_l^{2l+1}r^2}{2}  \ ,
\end{equation*}
the second one is:
\begin{equation*}
\frac{u_l^{(2)}}{u_l^{(0)}}[(l+1)r^lu_l^{(0)}-r^{l+1}u_l^{(0)'}]=\frac{2l+1}{
2l+3}\frac{a_l^{2l+1}r^2}{2}+\frac{(2l+1)^2}{(2l+3)(2l-1)}a_l^{4l+2}\Big(r^{1-2l
}+a_l^{2l+1}r^{-4l}\Big)  \ ,
\end{equation*}
and the last one is:
\begin{equation*}    
\frac{r^{l+1}}{u_l^{(0)}}\int_0^{r}{u_l^{(0)}}^2(r)dr=\int_0^{r}{u_l^{(0)}}
^2(r)dr+\frac{a_l^{2l+1}r^2}{2l+3}+a_l^{4l+2}r^{1-2l}\bigg(\frac{1}{2l+3}
-1\bigg)+a_l^{6l+3}r^{-4l}\bigg(\frac{1}{1-2l}+\frac{1}{2l+3}-1\bigg) \ .
\end{equation*}
Combining all terms yields
\begin{equation*}
    \frac{(2l+1)c_2}{2}=\frac{r^{2l+3}}{2l+3}-\frac{a_l^{2l+1}r^2}{2}-\frac{2l+1}{2l+3}\frac{a_l^{2l+1}r^2}{2}-\frac{(2l+1)^2}{(2l+3)(2l-1)}a_l^{4l+2}r^{1-2l}
\end{equation*}
\begin{equation*}
    -\frac{(2l+1)^2}{(2l+3)(2l-1)}a_l^{6l+3}r^{-4l}-\int_0^{r}{u_l^{(0)}}^2(r)dr-\frac{a_l^{2l+1}r^2}{2l+3}+a_l^{4l+2}r^{1-2l}\frac{2l+2}{2l+3}\times\frac{1-2l}{1-2l}
\end{equation*}
\begin{equation*}
    -a_l^{6l+3}r^{-4l}\frac{(2l+3)(2l-1)-(2l-1)+(2l+3)}{(2l+3)(1-2l)}
\end{equation*}
\begin{equation}
    =\frac{r^{2l+3}}{2l+3}-a_l^{2l+1}r^2+\frac{a_l^{4l+2}r^{1-2l}}{1-2l}-\int_0^{r}{u_l^{(0)}}^2(r)dr
\end{equation}
Finally, $c_2$ is given by
\begin{equation}
c_2=\frac{2}{2l+1}\bigg\{\frac{r^{2l+3}}{2l+3}-a_l^{2l+1}r^2+\frac{a_l^{4l+2}r^{1-2l}}{
1-2l}-\int_0^{r}{u_l^{(0)}}^2(r)dr\bigg\}  \ . 
\end{equation}

\section{Equivalence of the effective range expression with the results from Ref. \cite{lars}}
\label{appendix:d}
In order to prove that our expressions (Eq. (\ref{app.eq1})) and the one found by L. B. Madsen in Ref. \cite{lars} (Eq. (\ref{app.eq2})) are equivalent, we need to introduce the different definitions we have used. We point out that, in Ref. \cite{lars},  Eqs. (33) and (36) have a typo. Equation (36) is precisely the one we are comparing here (Eq. (\ref{app.eq2})). In those equations, the term containing a $1/R^{2l-1}$ only has the double factorial of $(2l-1)$ to the power 1, while it should have the double factorial squared. In this work, we compare our formulas with Eq. (36) of Ref. \cite{lars} but with double factorial squared.
\begin{equation}
    \begin{aligned}
 r_0^{\text{eff}}=\frac{2}{a_0^2}\int_0^{\infty}\Big[(r-a_0)^2-{u_0^{(0)}}
^2(r)\Big ] dr\\
   r_{l>0}^{\text{eff}}=\frac{2}{(2l+1)a_l^{2l+2}}\int_0^{\infty}\Big[r^{2l+2}
-2a_l^ { 2l+1 }r-{u_l^{(0)}}^2(r)\Big]dr
    \end{aligned}
    \label{app.eq1}
\end{equation}
\begin{equation}
r_l^{\text{eff},*}=\lim_{k\rightarrow0}2\bigg(\int_0^Rdr[v_{l,0}(r)v_{l,k}(r)-u_{l
, 0 } 
(r)u_{l,k}(r)]-\frac{1}{a_l^*}\frac{R^2}{(2l+1)}-\frac{[(2l-1)!!]^2}{(2l-1)R^{2l-1}} 
\bigg) \ .
\label{app.eq2}
\end{equation}

Before continuing, we define the following relations:
\begin{equation}
    A_l=\frac{1}{(2l+1)!!}=\frac{2^ll!}{(2l+1)!}\quad\text{and}\quad B_l=(2l-1)!!=\frac{(2l)!}{2^ll!} \ .
    \label{app.eq8}
\end{equation}

In our work, the cotangent of the phase shift is defined as
\begin{equation}
    k^{2l+1}\cot{\delta}\approx\frac{B_l}{A_l}\frac{1}{a_l^{2l}}\bigg[-\frac{1}{a_l}+\frac{1}{2}r_l^{\text{eff}}k^2\bigg] \ ,
    \label{app.eq3}
\end{equation}
while Madsen defines it as
\begin{equation}
    k^{2l+1}\cot{\delta}\approx-\frac{1}{a_l^*}+\frac{1}{2}r_l^{\text{eff},*}k^2 \ .
    \label{app.eq4}
\end{equation}
By comparing Eq. (\ref{app.eq3}) and Eq. (\ref{app.eq4}), we can relate our $a_l$ and $r_l^{\text{eff}}$ with Madsen's $a_l^*$ and $r_l^{\text{eff},*}$:
\begin{equation}
a_l^*=\frac{A_l}{B_l}a_l^{2l+1}\quad \text{and}\quad r_l^{\text{eff},*}=\frac{B_l}{A_l}\frac{r_l^{\text{eff}}}{a_l^{2l}}=\frac{a_lr_l^{\text{eff}}}{a_l^*}
\label{app.eq5}
\end{equation}

For the large distance behavior, we make use of
\begin{equation}
    u_l(r)=\frac{k^{-l}}{A_l}r\Big[j_l(kr)-n_l(kr)\tan{\delta}\Big] \ ,
\end{equation}
which at zero energy ($k=0$) becomes $r^{l+1}-a_l^{2l+1}r^{-l}$. Instead, Madsen uses
\begin{equation}
    u_{l,k}(r)=c_{l,k}\Big[F_l(kr)\cot{\delta}+G_l(kr)\Big]=rk^{l+1}\Big[j_l(kr)\cot{\delta}-n_l(kr)\Big] \ ,
\end{equation}
with $c_{l,k}=k^l$, $F_l(kr)=krj_l(kr)$ and $G_l(kr)=-krn_l(kr)$. At $k=0$, it becomes $-B_la_l^{-2l-1}[r^{l+1}-a_l^{2l+1}r^{-l}]$. The other function that Madsen uses is
\begin{equation}
    v_{l,k}(r)=c_{l,k}F_l(kr)\cot{\delta}=rk^{l+1}j_l(kr)\cot{\delta} \ ,
\end{equation}
which becomes  $-B_la_l^{-2l-1}r^{l+1}$ at zero energy. This means that Madsen is using the same functions that we use but with a multiplying factor that is $-B_la_l^{-2l-1}$.

Now we can address the fact that Eq. (\ref{app.eq1}) and Eq. (\ref{app.eq2}) are equivalent. Let's start by using the relation between effective ranges (Eq. (\ref{app.eq5})). 
\begin{equation}
    r_l^{\text{eff}}=\frac{a_l^*}{a_l}r_l^{\text{eff},*}=2\frac{a_l^*}{a_l}\bigg(\int_0^Rdr[v_{l,0}^2(r)-u_{l,0}^2(r)]-\frac{1}{a_l^*}\frac{R^2}{(2l+1)}-\frac{[(2l-1)!!]^2}{(2l-1)R^{2l-1}} 
\bigg)
\label{app.eq6}
\end{equation}
In Eq. (\ref{app.eq6}) we have already applied the limit $k\rightarrow0$. Now we will pass from the functions that Madsen uses to ours, and the factor $-B_la_l^{-2l-1}$ squared will be taken out as a common factor:
\begin{equation}
    r_l^{\text{eff}}=2\frac{B_l^2a_l^*}{a_l^{4l+3}}\bigg(\int_0^Rdr[r^{2l+2}-{u_l^{(0)}}^2(r)]-\frac{a_l^{4l+2}}{B_l^2a_l^*}\frac{R^2}{(2l+1)}-\frac{a_l^{4l+2}}{(2l-1)R^{2l-1}} 
\bigg)
\label{app.eq7}
\end{equation}
If we use the expressions of $A_l$, $B_l$ (Eq. (\ref{app.eq8})) and $a_l^*$ (Eq. (\ref{app.eq5})), we find that
\begin{equation}
    B_l^2a_l^*=\frac{a_l^{2l+1}}{2l+1} \ .
\end{equation}
With this, we can simplify the coefficients we have in Eq. (\ref{app.eq7}), obtaining
\begin{equation}
    r_l^{\text{eff}}=\frac{2}{(2l+1)a_l^{2l+2}}\bigg(\int_0^Rdr[r^{2l+2}-{u_l^{(0)}}^2(r)]-a_l^{2l+1}R^2+\frac{a_l^{4l+2}R^{1-2l}}{(1-2l)} 
\bigg)
\label{app.eq9}
\end{equation}
Equation (\ref{app.eq9}) is the same as Eq. (49) of Ref. \cite{Part-I}. If we now apply the limit $R\rightarrow\infty$, Eq. (\ref{app.eq9}) will give Eq. (\ref{app.eq1}). Therefore,  both results are equivalent.

\section{Effective range for diverging scattering length}
\label{appendix:c}
We intend to make $a_l$ explicit in our expressions as we want to take the 
limit $a_l\to\pm\infty$. To do so, we take into account the fact that we know 
how the wave function behaves at large distances. We split the wave function in 
two parts, and we impose these two parts to behave at $r\to\infty$, as we 
expect from the large-distance solution. The large-distance solution for the 
wave function and its derivative is 
\begin{equation}
\begin{aligned}
    u_l^{(0)}(r\rightarrow\infty)=r^{l+1}-a_l^{2l+1}r^{-l}\\
    {u'}_l^{(0)}(r\rightarrow\infty)=(l+1)r^{l}+la_l^{2l+1}r^{-l-1} \ .
\end{aligned}
\end{equation}
If we combine the wave function and its derivative with the proper 
coefficients, 
we can manage to obtain what we intend. After the split, the wave function can 
be expressed as
\begin{equation}
    u_l^{(0)}(r)=\frac{lu_l^{(0)}(r)+r{u'}_l^{(0)}(r)}{1+2l}-\frac{r{u'}_l^{(0)}(r)-(l+1)u_l^{(0)}(r)}{1+2l} \ .
\end{equation}
 The first part of the split gives $r^{l+1}$,
\begin{equation} 
\lim_{r\rightarrow\infty}\frac{lu_l^{(0)}(r)+r{u'}_l^{(0)}(r)}{1+2l}=\frac{lr^{ 
l+1}-la_l^{2l+1}r^{-l}+r(l+1)r^{l}+rla_l^{2l+1}r^{-l-1}}{1+2l}=r^{l+1} \ .
\end{equation}
The same for the second part giving $a_l^{2l+1}r^{-l}$,
\begin{equation}
\lim_{r\rightarrow\infty}\frac{r{u'}_l^{(0)}(r)-(l+1)u_l^{(0)}(r)}{1+2l}=\frac{
r(l+1)r^{l}+rla_l^{2l+1}r^{-l-1}-(l+1)r^{l+1}+(l+1)a_l^{2l+1}r^{-l}}{1+2l}=a_l^{
2l+1}r^{-l} \ . 
\end{equation}
As long as $a_l$ is non-zero, we can redefine the two parts as two functions,
\begin{equation}
\label{eq.deffh}
    \frac{lu_l^{(0)}(r)+r{u'}_l^{(0)}(r)}{1+2l}\equiv f_l(r)\quad;\quad \frac{r{u'}_l^{(0)}(r)-(l+1)u_l^{(0)}(r)}{1+2l}\equiv a_l^{2l+1}h_l(r) \ .
\end{equation}
Now, we want to know how these functions $f_l(r)$ and $h_l(r)$ behave. Hence, we need to know what differential equation they satisfy. Let's start with 
$f_l(r)$. First of all, we will express $f_l(r)$, its derivative ${f_l}'(r)$, 
and its second derivative ${f_l}''(r)$ in terms of $u_l^{(0)}(r)$ and 
${u'}_l^{(0)}(r)$. To this end, we will make use of the radial Schr\"odinger 
equation:
\begin{equation}
    \begin{aligned}
        f_l(r)=\frac{lu_l^{(0)}(r)+r{u'}_l^{(0)}(r)}{1+2l}\\     
{f_l}'(r)=\frac{(l+1){u'}_l^{(0)}(r)+r{u''}_l^{(0)}(r)}{1+2l}=\frac{(l+1){u'}_l^ 
{(0)}(r)+rU_l^{\text{eff}}(r)u_l^{(0)}(r)}{1+2l}\\     
{f_l}''(r)=\frac{(l+2){u''}_l^{(0)}(r)+r{u'''}_l^{(0)}(r)}{1+2l}=\frac{\Big[ 
(l+2)U_l^{\text{eff}}(r)+r{U'}_l^{\text{eff}}(r)\Big]u_l^{(0)}(r)+rU_l^{\text{
eff} } (r) { u' } _l^ { (0) } (r) 
}{1+2l}  \ ,
    \end{aligned}
\end{equation}
with $U_l^{\text{eff}}(r) = U(r) + l(l+1)/r^2$.   If we now apply 
$A{f_l}''(r)+B{f_l}'(r)+Cf_l(r)=0$, we  get
\begin{equation}
    \begin{aligned}
        \text{If}\quad U(r)=0\quad\Rightarrow\quad A=0\quad;\quad B=1\quad;\quad C=-\frac{l+1}{r}\\
         \text{Else}\quad\Rightarrow\quad A=1\quad;\quad 
B=-\bigg(\frac{2}{r}+\frac{U'(r)}{U(r)}\bigg)\quad;\quad 
C=\bigg(\frac{(l+1)}{r}\frac{U'(r)}{U(r)}+\frac{2+l-l^2}{r^2}-U(r)
\bigg)  \ .
    \end{aligned}
\end{equation}
Therefore, $f_l(r)$ satisfies the following differential equation:
\begin{equation}
    \begin{aligned}
        \text{If}\quad U(r)=0\quad\Rightarrow\quad {f_l}'(r)-\frac{l+1}{r}f_l(r)=0\\
         \text{Else}\quad\Rightarrow\quad 
{f_l}''(r)-\bigg(\frac{2}{r}+\frac{U'(r)}{U(r)}\bigg){f_l}'(r)+\bigg(\frac{(l+1)
}{r}\frac{U'(r)}{U(r)}+\frac{2+l-l^2}{r^2}-U(r)\bigg)
f_l(r)=0   \ .
    \end{aligned}
\end{equation}
The case $U(r)=0$ corresponds to the large-distance regime, thus giving 
$r^{l+1}$. For $h_l(r)$,
\begin{equation}
    \begin{aligned}
        a_l^{2l+1}h_l(r)=\frac{r{u'}_l^{(0)}(r)-(l+1)u_l^{(0)}(r)}{1+2l}\\     
a_l^{2l+1}{h_l}'(r)=\frac{r{u''}_l^{(0)}(r)-l{u'}_l^{(0)}(r)}{1+2l}=\frac{rU_l^{ 
\text{eff}}(r)u_l^{(0)}(r)-l{u'}_l^{(0)}(r)}{1+2l}\\         
a_l^{2l+1}{h_l}''(r)=\frac{(1-l){u''}_l^{(0)}(r)+r{u'''}_l^{(0)}(r)}{1+2l}=\frac 
{\Big[(1-l)U_l^{\text{eff}}(r)+r{U'}_l^{\text{eff}}(r)\Big]u_l^{(0)}(r)+rU_l^{\text{eff
} } (r) { u' } _l^ { 
(0)}(r)}
{1+2l} \ .
    \end{aligned}
\end{equation}
If we now apply $A{h_l}''(r)+B{h_l}'(r)+Ch_l(r)=0$, we  get
\begin{equation}
    \begin{aligned}
        \text{If}\quad U(r)=0\quad\Rightarrow\quad A=0\quad;\quad B=1\quad;\quad C=\frac{l}{r}\\
         \text{Else}\quad\Rightarrow\quad A=1\quad;\quad 
B=-\bigg(\frac{2}{r}+\frac{U'(r)}{U(r)}\bigg)\quad;\quad 
C=-\bigg(\frac{l}{r}\frac{U'(r)}{U(r)}+\frac{3l+l^2}{r^2}+
U(r)\bigg)  \ .
    \end{aligned}
\end{equation}
Therefore, $h_l(r)$ satisfies the following differential equation,
\begin{equation}
    \begin{aligned}
        \text{If}\quad U(r)=0\quad\Rightarrow\quad {h_l}'(r)+\frac{l}{r}h_l(r)=0\\
         \text{Else}\quad\Rightarrow\quad 
{h_l}''(r)-\bigg(\frac{2}{r}+\frac{U'(r)}{U(r)}\bigg){h_l}'(r)-\bigg(\frac{l}{r}
\frac{U'(r)}{U(r)}+\frac{3l+l^2}{r^2}+U(r)\bigg)
h_l(r)=0   \ .
    \end{aligned}
\end{equation}
Again, the case $U(r)=0$ corresponds to the large-distance regime, giving $r^{-l}$. Both $f_l(r)$ and $h_l(r)$ satisfy the same boundary condition as $u_l^{(0)}(r)$, which is $f_l(r=0)=0$ and $h_l(r=0)=0$, both functions at the origin must be zero. This can be seen if we look at Eq. (\ref{eq.deffh}) and we set $r=0$. For completeness, as we have proved above, the large-distance solutions are $f_l(r \to \infty)=r^{l+1}$ and $h_l(r \to \infty)=r^{-l}$. With this, the wave function can be rewritten as $u_l^{(0)}(r)=f_l(r)-a_l^{2l+1}h_l(r)$ as long as $a_l$ is not zero. If we apply this transformation into the effective range, we obtain
\begin{equation}
r_l^{\text{eff}}=\frac{2}{(2l+1)a_l^{2l+2}}\bigg\{\int_0^{\infty}\Big[r^{2l+2}-f_l^2(r)\Big]dr-2a_l^{2l+1}\int_0^{\infty}\Big[r-f_l(r)h_l(r)\Big]+a_l^{
4l+2}\int_0^{\infty}\Big[r^{-2l}\delta_{l,0}-h_l^2(r)\Big]dr\bigg\} \ .
\end{equation}
The above integrals are finite. This can be seen by knowing that $f_l^2(0)$, 
$f_l(0)h_l(0)$ and $h_l^2(0)$ are zero, and that 
$\lim_{r\rightarrow\infty}\Big[r^{2l+2}-f_l^2(r)\Big]=0$, 
$\lim_{r\rightarrow\infty}\Big[r-f_l(r)h_l(r)\Big]=0$, and 
$\lim_{r\rightarrow\infty}\Big[r^{-2l}\delta_{l,0}-h_l^2(r)\Big]=0$.


\bibliography{apssamp2}